\documentclass[useAMS,usenatbib]{mn2e}
\usepackage{graphics}
\usepackage{subfigure}
\usepackage{graphicx}
\usepackage{amssymb}
\usepackage{hyperref}
\title[Effects of AGN feedback on $\Lambda$CDM galaxies]{Effects of AGN feedback on $\Lambda$CDM galaxies} 
\author[Claudia del P. Lagos, Sof\'ia A. Cora, Nelson D.
Padilla]{Claudia del P. Lagos$^{1}$, Sof\'ia A. Cora$^{2,3}$, Nelson D.
Padilla$^{1}$\\
$^{1}$Departamento Astronom\'ia y Astrof\'isica, Pontificia Universidad
Cat\'olica de Chile, Av. Vicu\~na Mackenna 4860, Stgo., Chile\\
$^{2}$ Facultad de Ciencias Astron\'omicas y Geof\'isicas de la Universidad Nacional
de La Plata, and Instituto de Astrof\'isica de La Plata \\(CCT La Plata, CONICET, UNLP), Observatorio Astron\'omico, Paseo del Bosque S/N, 1900 La Plata, Argentina\\
$^{3}$Consejo Nacional de Investigaciones Cient\'{\i}ficas y T\'ecnicas,
Rivadavia 1917, Buenos Aires, Argentina\\
}
\begin{document}

\date{Accepted ???. Received ???; in original form
2008 April 8}

\pagerange{\pageref{firstpage}--\pageref{lastpage}} \pubyear{2008}

\maketitle

\label{firstpage}

\begin{abstract}
We study the effects of Active Galactic Nuclei (AGN) feedback on the formation and evolution of galaxies by
using a combination of a cosmological {\em N}-body simulation of the concordance $\Lambda$
cold dark matter ($\Lambda$CDM) paradigm and a semi-analytic model of galaxy formation.
This model is an improved version of the one described by \citet{Cora06}, 
which now considers the growth of black holes (BHs) as driven by (i) gas
accretion during merger-driven starbursts and mergers with other BHs,
(ii) accretion during starbursts triggered by disc instabilities, and (iii) accretion of 
gas cooled from quasi-hydrostatic hot gas haloes.  It is assumed that
feedback from AGN operates in the later
case.  The model has been calibrated in order to reproduce observational
correlations between BH mass and 
mass, velocity dispersion, and absolute magnitudes of the galaxy bulge.
AGN feedback has a strong impact on reducing or even suppressing gas cooling, an effect
 that becomes important at lower redshifts.  This phenomenon helps 
to reproduce the observed galaxy luminosity function (LF)
in the optical and near IR bands at $z=0$, and the cosmic star formation rate 
and stellar mass functions
over a wide redshift range ($0 \la z \la 5$). 
It also allows to have a population of massive galaxies
already in place at $z\ga 1$, which are mostly early-type and have 
older and redder stellar populations than lower mass galaxies,
reproducing the observed bimodality in the galaxy colour distribution, and the morphological fractions.
The evolution of the optical QSO LF is also reproduced,
provided that the presence of a significant fraction of obscured QSOs 
is assumed.
We explore the effects of AGN feedback during starbursts finding that, 
in order to obtain a good agreement with observations, 
these need to be strong enough to expell the reheated gas away from the
galaxy halo.
We also test new, recent prescriptions for dynamical 
friction time-scales, and find that they produce  
an earlier formation of elliptical galaxies, and
a larger amount of disc instabilites, which compensate the change in the 
merger frequency such that
the properties of $z=0$ galaxies remain almost unaffected.
\end{abstract}


\section{Introduction}\label{Introsec}

Over the last decade, there has been a phenomenal increase in the range
and quantity of data on quasi-stellar objects (QSOs) or `quasars' 
and galaxies with active galactic nuclei (AGN),
which are the seats of an intense production of energy. They 
were singled out from the general galaxy population by their peculiar colours, morphology and/or strong 
emission in the radio, X-ray or infrared bands. 
The conventional view regards QSOs and AGN
as different manifestations of the same phenomenon.
Observations of the brightest Seyfert
galaxies, one of the 
types in which AGN are classified,
show events of sudden bursts of star formation activity (starbursts) 
that occurred 1-2 Gyrs ago; these events
have been associated, by some authors, 
with galaxy mergers
(\citealt{kauffman03}; \citealt{Sanders96}). Recent results by
\citet{Feain07} show direct evidence of the QSO-star formation connection 
where $70\%$ of 
the quasar
radio emission comes from star
formation activity
of the host galaxy.

In a pioneering work by \citet{Hoyle63}, it was proposed that the energy
from AGN was of gravitational origin derived from the
collapse of very massive objects under their own strong gravitational fields.
These early ideas
found a modified expression in the paradigm of black hole accretion discs
a few years later \citep{LyndenBell69}.
In more recent times, an important connection between AGN and galaxy evolution has been deduced
from the correlation between black holes, thought to be the engines of AGN,  
and properties of the host galaxy such as
the bulge mass \citep{Haring04}, the luminosity in the $K$-, $B$- 
and $V$-bands, 
the stellar mass
\citep{marconi03}, and the velocity dispersion of the stars in the bulge
\citep{ferrarese00}. The existence of a galactic bulge has been thought to be
related to merger events, which led to speculate that mergers may drive the
formation and evolution of black holes.
Furthermore, \citet{boyle98} show a correlation
between the evolution of the global stellar formation rate and the luminosity
density of optically selected quasars. This could also indicate that black 
holes are related to merger events and, therefore, to starburst phenomena as 
many authors
have speculated (e.g. \citealt{Malbon07}; \citealt{Croton06};
\citealt{Bower06}; \citealt{Sijacki07}).

The alleged link between AGN and star formation activity brings questions 
related to the effect of this nuclear activity on the evolution of their host 
galaxies
or galaxy clusters.  
Many recent observational results have began to answer some of these
questions complementing our picture of galaxy formation and evolution. For
example, \citet{Schawinski07} find a population of early type galaxies
lying outside
the red sequence in the colour-magnitude diagram; this red population is comprised by star forming and
Seyfert galaxies. From these data, they identify an evolutionary sequence from
star forming (blue population) to quiescent galaxies (red population). This
transition is explained by the effect of an AGN phase occurring roughly 
$0.5$~Gyr after
the starburst, which suppresses star formation.
\citet{Reuland07} 
have been able to confirm this result at high redshifts.

Direct observational evidence of AGN activity is provided by radio galaxies and
quasars containing jets that 
transport energy from the AGN to the surrounding medium.
The effects of jets issuing from AGN have been observed 
in many different wavelengths.
Radio sources 
show large-scale outflows of significant gas masses which are generated by
radio jets 
(\citealt{Nesvadba06}; \citealt{Nesvadba07}; \citealt{Temi07}). 
Theoretical studies of jets indicate that the time-scales
for transient activity (e.g. restoration of equilibrium, buoyant transport in
the hot gas) are consistent with a release of feedback energy from a central 
black hole
\citep{Temi07}.
The dichotomy in the properties of low- and
high-excitation radio galaxies has been interpreted as an additional
contribution from cold gas acting in high-excitation sources that forms the 
cold disc and torus around
a central engine \citep{Evans07}. 
Observations in X-rays
indicate 
the presence of shocks, bubbles and sound waves
thought to be driven by nuclear activity and to be
responsible for heating
the cluster core, as in the Perseus cluster (\citealt{Sanders07}, and references
therein).

All these results illustrate the key role of active galactic nuclei 
on galaxy evolution. 
As AGN have become a crucial ingredient of the galaxy evolution process, 
many theoretical works have
proposed different possible scenarios
for BH growth and the associated AGN feedback.
These works include
pure analytic approximations 
(e.g. \citealt{Efstathiou88}; \citealt{Percival99}; \citealt{churazov02}; \citealt{Hopkins07}),
semi-analytic models of galaxy formation

(e.g.  \citealt{Kauffmann00}; \citealt{Cattaneo01}; \citealt{Enoki03};
\citealt{Granato04}; \citealt{Cattaneo05}; \citealt{Menci06}; \citealt{Cattaneo06}; 
\citealt{Croton06};
\citealt{Bower06}; \citealt{Malbon07}; \citealt{Marulli08}), and
fully self-consistent numerical models
(\citealt{Springel05a}; \citealt{Sijacki07}). 
The application of a semi-analytic model of galaxy formation on
a cosmological numerical simulation
has demonstrated to be
a very appropriate tool for the study of
phenomena that drive the formation and evolution of galaxies within the 
hierarchical
clustering scenario. 
This is the method used to investigate the effects of AGN feedback in the
present work.

Our semi-analytic model is based on the one
described by \citet{Cora06}, to
which we refer to as 
SAG1 (acronym for `Semi-Analytic Galaxies' version 1).
This model follows the formation and evolution
of galaxies including  
gas cooling, star formation, SNe explosions and mergers, as
well as a detailed implementation of
metal enrichment of stars and interstellar
and intergalactic 
medium. 
SAG1, as well as many other semi-analytic models, reproduce to a 
reasonable degree
observational results such as gas fractions, luminosity functions
(LF), 
the Tully-Fisher relation, and galaxy colours.  
However, 
some inconsistencies still persist, 
in particular, the existence of a blue massive galaxy
population and an excess
of bright galaxies in the LF compared with observations in the $b_{\rm J}$- and $K$-band
(e.g. \citealt{Springel01}; \citealt{Cole00}; \citealt{Baugh04}). 
These problems are in part due to the fact that
the mass function of host dark-matter haloes does not show a sharp cut-off at
masses corresponding to $L^*$ galaxies in the model. Therefore,
models that populate these haloes tend to produce too many bright galaxies
\citep{Baugh06}.
The implementation of an artificial supression of gas cooling made in
SAG1,
that affects  
haloes characterized by a virial velocity higher
than a certain threshold, 
was a simple solution to overcome, at least
partially, the inconsistencies mentioned above.
These problems
call for the inclusion of additional processes in order
to reconcile, to a further degree, model results with observations.
The physical process thought to be responsible for the control
of the amount of gas that can cool is AGN feedback,
which is
triggered by
gas accretion events onto
central supermassive black holes.
Recent works have made progress in this sense, demonstrating
that AGN feedback
is crucial to reduce gas cooling in large haloes,
thus preventing them from forming stars at late times
and being able to 
explain the exponential cut-off in the
the bright-end of the galaxy LF
(e.g. \citealt{Cattaneo06}; \citealt{Menci06}; \citealt{Croton06}; \citealt{Bower06}; \citealt{Malbon07}).
Following 
the prescriptions implemented in these works, 
we replace the simple procedure of stopping gas cooling
in haloes with high virial velocity, 
as it is implemented in SAG1, by a more physical modelling of
AGN feedback in galaxies, giving raise to the version SAG2.  

The currently implemented models of AGN feedback
only prevents significant gas cooling in large
haloes. The so called {\em radio mode} feedback (see for instance,
\citealt{Croton06})
is assumed to produce low energy `radio' activity.
Recent observational results by \citet{gastadello07} on the
galaxy cluster AWM 4 cannot be explained if AGN activity 
only produces this mode of feedback,
since it presents an active radio source
which is not accompanied by a X-ray temperature gradient; 
this might indicate
the presence of AGN feedback
during high accretion rates 
originated during mergers and disc instabilities, which also feed
the BH growth.
We explore this possibility within the 
$\Lambda$ cold dark matter ($\Lambda$CDM)
framework using SAG2. 

Our 
model SAG2 is also used to investigate the impact
of
new theoretical merger time-scales presented in two parallel
studies carried out by \citet{jiang07} and 
Boylan-Kolchin, Ma \& Quataert (2007).
These works show that the
equation for merger time-scales given by \citet{lacey93} systematically underestimates
the time-scales of minor mergers and overestimates those of major mergers.

This work is organised as follows. 
Section \ref{Modelsec} presents a description of the
cosmological simulation used and the improved semi-analytic
model SAG2, 
summarising first the original model SAG1 and then giving a detailed
description of the AGN feedback model introduced. 
Section \ref{resultsec} describes the 
behaviour
of BH and QSO 
properties in the model, and presents the analysis of 
the effects of AGN feedback on galaxy properties. 
Further improvements made to SAG2,
including the effects of 
starburst AGN feedback, and new
prescriptions for merger time-scales, are presented in Section \ref{Furthersec}.
Finally, Section \ref{finalsec} gives the main conclusions of this work.

\section{A hybrid model of galaxy formation and evolution}\label{Modelsec}
                                                                                
We study the formation and evolution of
galaxies by applying a numerical technique which combines 
a cosmological {\em N}-Body
simulation of the concordance $\Lambda$CDM universe
and a semi-analytic model of galaxy formation.  In this hybrid model,
the outputs of the cosmological
simulation are used to construct merger histories of dark matter haloes which
are used by the semi-analytic code to generate the galaxy population. 
The main advantage of semi-analytic codes is that they allow to 
reach a larger
dynamic range than fully self-consistent simulations, at a far smaller
computational cost.

The semi-analytic model used here is based on that
described by \citet{Cora06} (SAG1).
The physical processes considered in this model are cooling of hot gas as 
a result of 
radiative losses, star formation, feedback from supernova explosions
and galaxy mergers, modelled as in
\citet{Springel01}. The model
also tracks the circulation of metals between the different
baryonic components, that is, cold gas, hot diffuse gas, and stars,
following the model described by 
\citet{DeLucia04}, 
with the 
additional advantage
of including the mass evolution of different chemical elements.

In the present work, SAG1 is improved
by including feedback from AGN as a replacement 
of the artificial suppression of gas
cooling in massive haloes, thus leading to the version SAG2. 
In the following subsections, 
we present the main characteristics of the {\em N}-Body simulation
used,
and briefly describe the physical processes
already included in SAG1.
In the original version of the code, 
only major mergers
trigger starbursts; we now include two other mechanisms responsible
of this kind of events, minor mergers and disc instabilities.
Then, we present the model of black hole growth
and the associated AGN feedback 
implemented in the new version of the code.

\subsection{$\Lambda$CDM Cosmological simulation} \label{SimuLambda}

We use a cosmological simulation of the concordance 
$\Lambda$CDM cosmology in a periodic box of $60 \, h^{-1}$~Mpc. This box is
large enough to allow us to make a suitable analysis of the luminosity function 
of galaxies and quasars, and of the galaxy mass function.
The simulation contains
$16,777,216$ dark matter particles with a mass resolution of $1.001 \times
10^{9}\,h^{-1}\,M_{\odot}$. 
More than $54,000$ dark matter haloes 
have been
identified,
with the largest one having a mass of $5.36 \times 10^{14}\,h^{-1}\,M_{\odot}$.

The simulation parameters are consistent with the results of 
WMAP data
\citep{spergel03}, that is, $\Omega_{\rm m}=\Omega_{\rm DM}+\Omega_{\rm
baryons}=0.28$ (with a baryon fraction of $0.16$), $\Omega_{\Lambda}=0.72$ and
$\sigma_{8}=0.9$. The Hubble constant is
$H_{0}=100 \, h\, {\rm Mpc}^{-1}$, with $h=0.72$.  The gravitational softening
length, $\epsilon$, is $3.0 \,h^{-1}$~kpc.  
The simulation starts from a redshift
$z=48$ and is run using the public version of the {\small GADGET-2} code 
 \citep{Springel05b}. 

Dark haloes are first identified as virialized particle groups by a
friends-of-friends (FOF) algorithm.  A SUBFIND algorithm \citep{Springel01} is
then applied to these groups in order to find self-bound dark matter
substructures.  Merger trees are then constructed from these dark matter haloes
and their embedded substructures.

\subsection[]{Semi-analytic model SAG1}\label{sec_SAM}

We briefly comment the processes of gas cooling, star formation, 
supernova feedback and metal production included in the model 
SAG1; we refer the reader to \citet{Cora06} for a more detailed description.

We assume that hot gas always fills the dark matter haloes
following an isothermal distribution.
Initially, 
the hot gas mass is given by the baryon fraction of
the virial mass of the dark matter halo. It is subsequently modified
by the amount of gas that has cooled
and the stellar mass formed in the galaxies contained within the halo.  
The mass
of hot gas that cools in each dark matter halo 
in the underlying cosmological simulation
is given by the cooling rate

\begin{equation}
\frac{dM_{\rm cool}}{dt}=4 \pi \rho_{\rm g} r_{\rm cool}^{2} \frac{dr_{\rm
cool}}{dt},
\label{cooledrate}
\end{equation}

\noindent where $\rho_{\rm g}$ is the density profile of an isothermal 
sphere, and $r_{\rm cool}$ is
the cooling radius. The local cooling time is defined as the ratio between the
specific thermal energy content of the gas and the cooling rate per unit volume,
$\Lambda(T,Z)$, which depends of the metallicity \citep{suth93}, and 
the temperature
$T=35.9(V_{\rm vir}/ {\rm km} \, {\rm s}^{-1})$ of the halo,
being $V_{\rm vir}$ its virial velocity.
This process only operates when the galaxy
is the central galaxy of a given halo.
The cooled gas contributes to the formation of the galaxy disc.
The cold gas mass of each galaxy is involved in the star
formation process.
The star formation
rate is given by,

\begin{equation}
\frac{dM_{\star}}{dt}= \frac{\alpha M_{\rm ColdGas}}{t_{\rm dyn}^{\rm gal}},
\label{sfr}
\end{equation}

\noindent where $\alpha$ is a parameter which regulates the efficiency of star
formation, $M_{\rm ColdGas}$ is the cold gas mass, and $t_{\rm dyn}^{\rm gal}=0.1 R_{\rm vir}/V_{\rm vir}$ is the dynamical
time of the galaxy, being 
$R_{\rm vir}$ the virial radius of the halo.
Here $\alpha=\alpha_{0}(V_{\rm vir}/220 \, {\rm km}\,{\rm s}^{-1})^{n}$, where $\alpha_{0}$ and $n$ are free
parameters set to 0.1 and 2.2, respectively. Each star formation event generates a stellar mass $\Delta
M_{\star}$, which leads to a number $\eta_{\rm CC}$ of core collapse supernovae
(SNe CC). 
This quantity depends on the initial mass function (IMF) adopted, which in this case is a Salpeter IMF
normalized between $0.1$ and $100 \, M_{\odot}$; therefore,  $\eta_{\rm CC}=6.3 
\times 10^{-3}$ $M_{\odot}^{-1}$. SNe CC include types Ib/c and II. 
The energy $E_{\rm SNCC}$ released by
each SN CC (1.2 $\times$ $10^{51} \, {\rm erg}\,{\rm s}^{-1}$)
is assumed to reheat the cold gas of a galaxy inducing galactic outflows that 
transfer
cold gas to the hot phase by the amount 
\begin{equation}
\Delta M_{\rm reheated}=\frac{4}{3} \epsilon \frac{\eta_{\rm CC} E_{\rm
SNCC}}{V_{\rm Vir}^{2}}
\Delta M_{\star},
\label{sncc}
\end{equation}

\noindent where $\epsilon$ is a dimensionless parameter which regulates the
efficiency of the feedback by SNe CC, and takes a value of $\epsilon=0.1$.
The reheated gas is kept within its host dark matter halo 
(`retention model', see \citealt{DeLucia04} for other possibilities).

We set an initial primordial abundance
of the baryonic components consisting of 76\%
hydrogen and 24\% helium. 
They
become chemically enriched 
as star formation and metal production take place.
Stars contaminate the cold and hot gas as a result of mass losses during their
evolution.
A fraction $f_{\rm ejec}$ of the mass ejected by galaxies 
is transferred directly to the hot phase instead of being first 
incorporated to the cold gas. 
We adopt here $f_{\rm ejec}=0.5$, a value that 
allows to 
obtain 
an evolution of the mass-metallicity relation
 consistent with  
observational data \citep{erb06}, 
resulting in 
a more gradual
contamination of the cold gas associated to each galaxy.
The transfer of reheated enriched cold gas also contributes to the
chemical enrichment of the hot phase.

Since the cooling rate depends on the metallicity,
this chemical enrichment has a strong influence on the amount of hot gas 
that can
cool.  This process in turn influences the star formation activity which is 
ultimately
responsible for the chemical pollution. Thus, new stars 
formed from the polluted cold gas become more chemically enriched.
Stars not only form in a quiescent way from the reservoir of cold gas 
(Eq.~\ref{sfr}) 
but also 
from starbursts
triggered by mergers and disc instabilities, 
as we will describe in Section~\ref{SBsec}.
The chemical model implemented in our semi-analytic code considers 
that metals are
provided by
three kinds of sources:
low and intermediate mass stars, 
SNe CC, and type Ia supernovae (SNe Ia); the first group of
sources yields metals through mass losses and stellar winds
(details are given in \citealt{Cora06}).

With respect to the spectro-photometric properties of galaxies, we consider
evolutionary synthesis models that depend on the metallicity of the cold gas
from which stars are formed, estimated for a Salpeter IMF 
\citep{bruzual03}.
All magnitudes and colours include the effects of dust extinction following
the implementation by \citet{DeLucia04}, which is based on previous
works (e.g. \citealt{kauffman99a}).  

It is important to remark that SAG1 suppresses gas cooling in rapidly
rotating dark haloes ($\rm V_{\rm Vir} > 350 \,{\rm km}\, {\rm s}^{-1}$).
This restriction is set with the aim of avoiding very massive and
luminous galaxies,
thus achieving a  
better agreement with observations of the bright end of the luminosity function.
The reduction or suppression of the cooling flow can be naturally
produced by AGN feedback (\citealt{Croton06}, \citealt{Bower06}).
This important physical process is now implemented in the
semi-analytic code as we describe in Section~\ref{BG_AGN}.

\subsection{Starbursts in the model: Galaxy mergers and Disc
Instabilities}\label{SBsec}

Our new version of the semi-analytic model produces starbursts in three
different ways: major mergers, minor mergers and disc
instabilities.
Starbursts occurring in the last two processes are new ingredients added to
the original version SAG1.
All the cold gas contained in the galaxies involved in any of these three 
processes 
is completely consumed in a sudden burst of star formation.

In a hierarchical scenario of structure formation, mergers of galaxies
are a natural consequence of the mergers of
dark matter haloes in which they reside, and 
play an important role in determining the mass
and morphology of galaxies. 
In the subhalo scheme arising from the identification of 
dark matter substructures, 
we distinguish three types of galaxies when
tracking galaxy formation.  The largest
subhalo in a FOF group hosts the `central galaxy' of the group; its
position is given by the most bound particle in that subhalo. Central
galaxies of other smaller subhaloes contained in a FOF group
are referred to as `halo galaxies'.  The (sub)haloes of these 
satellite
galaxies
are still intact after falling into larger structures.  The third
group of galaxies comprises `satellite' galaxies that
arise when two subhaloes merge and the galaxy of the
smaller one becomes a satellite of the remnant subhalo.  
In SAG1, satellite
galaxies are assumed to merge on a dynamical time-scale with the
central galaxy of the largest subhalo of the FOF group they reside in.
It is assumed that the satellite galaxy has a circular orbit with a
velocity given by the virial velocity of the parent halo and decays 
to the corresponding central galaxy
through dynamical friction \citep{binney87}, given by
\begin{equation} 
T_{\rm friction}= \frac{1}{2} \frac{f(\epsilon)}{\rm C} \frac{V_{\rm c}r_{\rm c}^{2}}{G
M^{\rm sat}
{\rm ln} \Lambda},
\label{dynfriction}
\end{equation}
where M$_{\rm sat}$ is the satellite mass orbiting at a radius
r$_{\rm c}$ inside an isothermal halo of circular velocity V$_{\rm c}$. 
The function
$f(\epsilon)$ describes the dependence on the eccentricity of the satellite 
orbit 
(see \citealt{Springel01} for a detailed description), and $C=0.43$. The
function ${\rm ln} \Lambda = (1 + M^{\rm  central}/M^{\rm sat})$ 
is the Coulomb logarithm.

Mergers are classified according to the ratio between the mass of the
accreted satellite galaxy and the mass of the central galaxy,
$f_{\rm merge}=M^{\rm sat}/M^{\rm central}$.
A major merger occurs when
$f_{\rm merge}>0.3$; in this case
all stars present are rearranged into a spheroid and all the cold gas in the
merging galaxies is assumed to undergo a 
starburst.
The stars thus
produced are added to the 
spheroid
(bulge).
Major mergers constitute a mechanism
to produce elliptical galaxies.
Subsequent accretion of cooling gas is funneled into the central galaxy
generating a disc.

On the other hand, 
the merger is classified as
a minor merger 
when one of the two galaxies has a small mass compared 
to the other, that is, $f_{\rm merge}<0.3$. 
Following \citet{Malbon07}, we assume that 
the presence of a starburst in a minor merger depends on the 
gas mass fraction of the disc of the central galaxy,
$f_{\rm ColdGas}^{\rm central}=M_{\rm ColdGas}^{\rm central}/M_{\rm disc}^{\rm
central}$, where
$M_{\rm disc}=M_{\rm Stellar}-M_{\rm Bulge}+M_{\rm ColdGas}$.
In this last relation, $M_{\rm Stellar}$ is the total stellar mass
of the galaxy,
$M_{\rm Bulge}$ is the mass of the bulge (formed only by stars), and
$M_{\rm ColdGas}$ is the cold gas content of the galaxy.
If $f_{\rm ColdGas}^{\rm central}>f_{\rm gas,\rm burst}$, 
where $f_{\rm gas,burst}$
is a free parameter of the model, the minor
merger will produce a starburst 
as a result of
the perturbation introduced by the
merging satellite galaxy which drives all the cold gas from both galaxies 
into the spheroid, where it is completely
transformed into stars.
Otherwise, if 
$f_{\rm ColdGas}^{\rm central}<f_{\rm gas,burst}$
then no burst occurs. We adopt 
a threshold $f_{\rm gas,burst}=0.6$. Furthermore,
if the 
satellite
is much less massive than the 
central galaxy
($f_{\rm merge}<f_{\rm burst}$, with $f_{\rm burst}=0.05$, as in 
\citealt{Malbon07}), the 
central
stellar disc remains unchanged and the 
accreted stars
are added to the spheroid; in this case, there is no starburst, 
regardless of the
amount of gas in the 
central
disc, which already contains the contribution of
the gas of the satellite.

Starbursts are also triggered by disc instabilities.  
When a galaxy disc
is sufficiently massive that its self-gravity is dominant, it becomes
unstable to small perturbations by minor satellites or dark matter substructures
(Mo, Mao \& White 1998; \citealt{Cole00}). 
We follow the stability criterion from
\citet{Cole00}, given by
\begin{equation} \epsilon =
\frac{V_{\rm max}}{(G M_{\rm disc} / r_{\rm disc})^{1/2}},
\label{eqDisc}
\end{equation}
where $V_{\rm max}$ is the maximum circular velocity of the disc, 
and $M_{\rm disc}$ is the disc mass.  We assume that
$r_{\rm disc}$ is the disc-scale radius (following \citealt{Mo98}).  When $\epsilon < \epsilon_{\rm disc}$ 
the disc becomes unstable; in the presence of a perturbation
all the stars and cold gas in the
disc are transferred to the bulge. In this case, all the gas present in this 
modified bulge is assumed to be consumed in a starburst. 
After this process, discs are required to stabilize before a new
starburst by disc instability can be triggered.
We adopt $\epsilon_{\rm disc}=1.1$ taking into 
account different observational contraints, as we describe 
subsequently along the paper.
This value lies within the limits given by \citet{Mo98} 
(ranging from $0.8$ to $1.2$)
for dark matter haloes with NFW profiles.
This mechanism also generates elliptical galaxies.

\begin{figure}
\begin{center}
\includegraphics[width=0.48\textwidth]{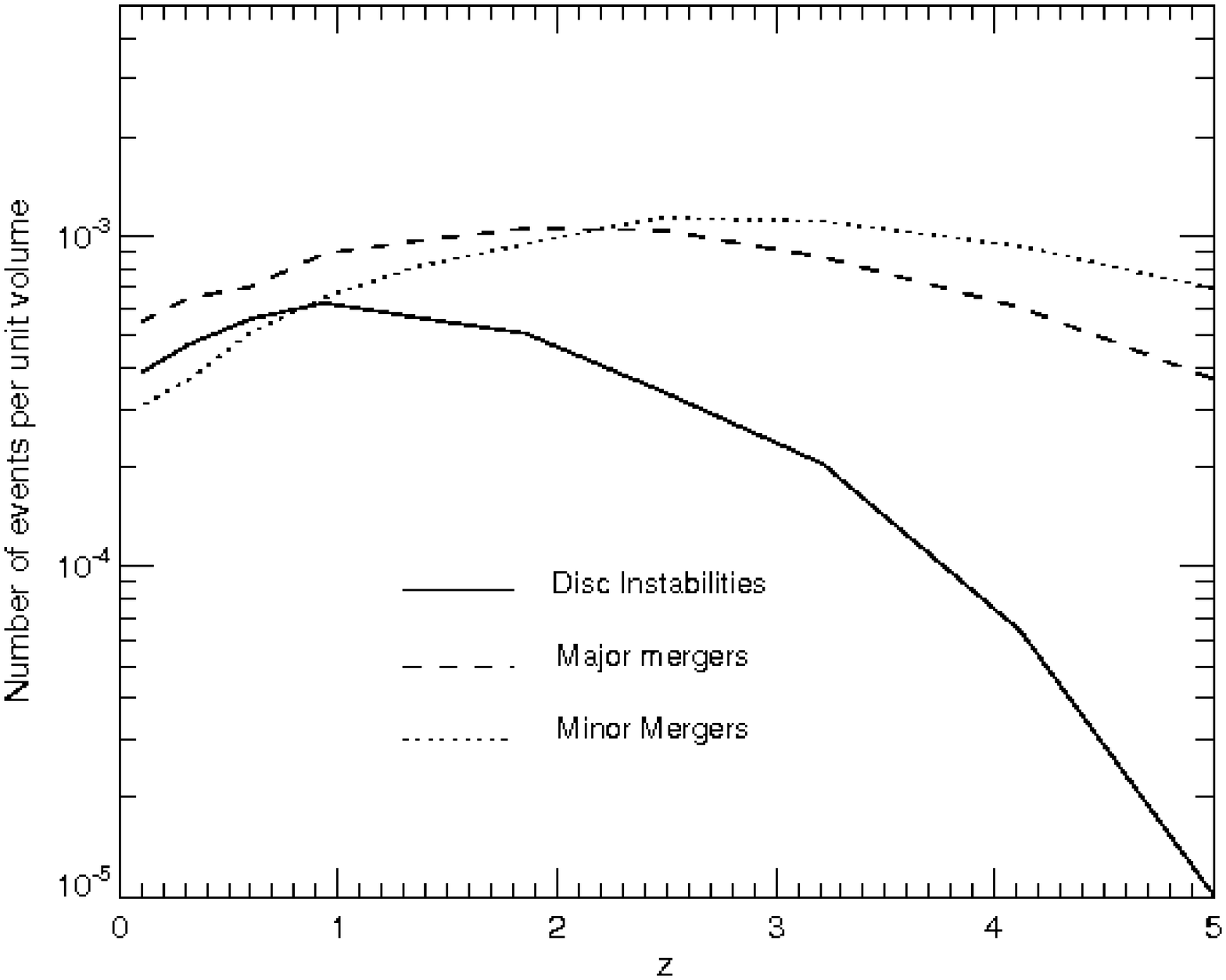}
\includegraphics[width=0.48\textwidth]{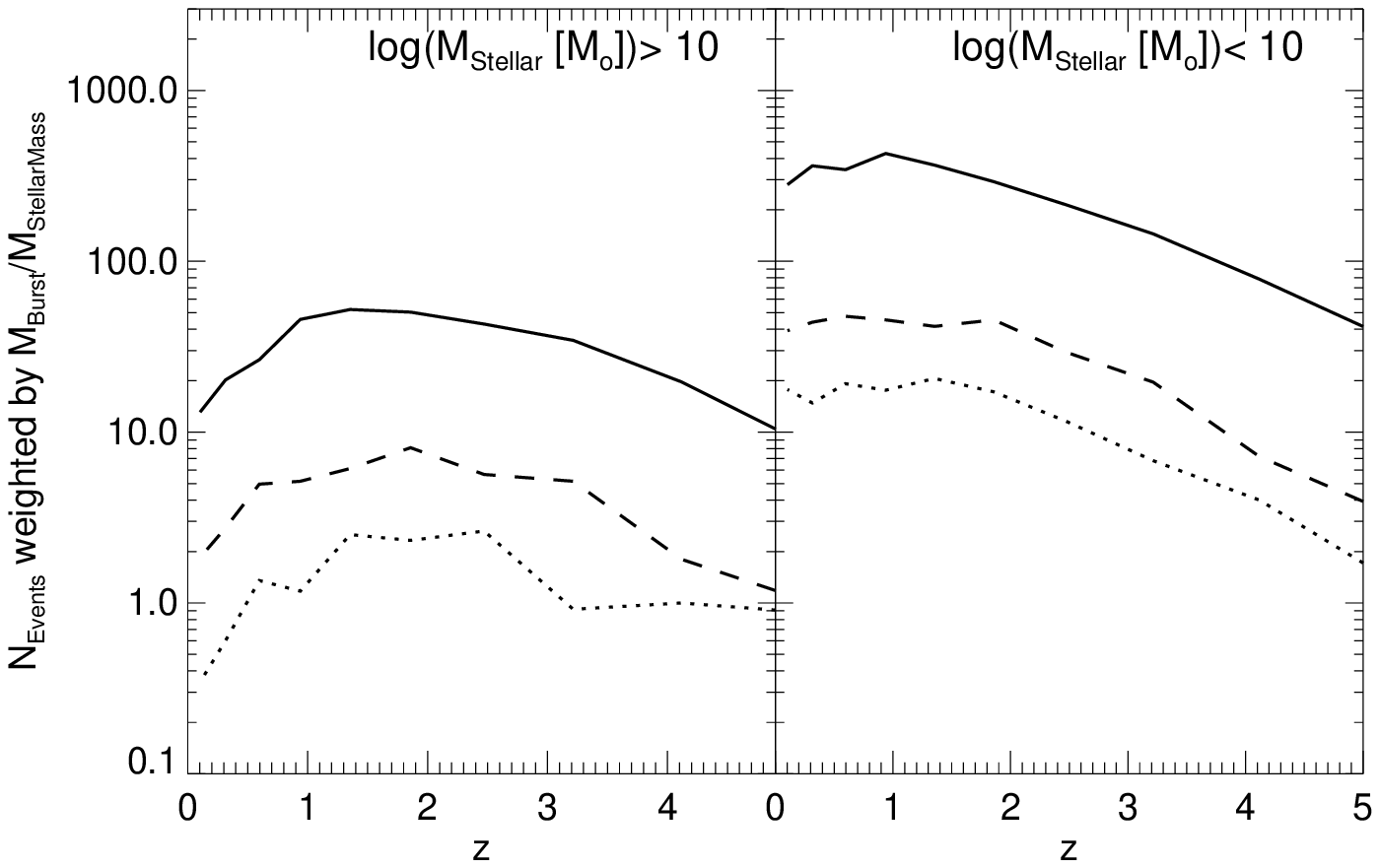}
\caption{{\em Upper panel}: Evolution with redshift of the number density
of events that trigger starbursts in our model: major mergers (dashed lines), minor mergers 
with $f_{\rm gas}>f_{\rm gas,burst}$ (dotted lines), and disc
instabilities (solid lines). {\em Lower panels}: 
Evolution with redshift of the
number of events weighted by 
the ratio between the mass of stars produced
by the burst 
and
the stellar mass present at the moment of the burst 
for high (lower left) and low (lower right) stellar mass galaxies.  Lines are as in the upper panel.}
\label{merTotDM}
\end{center}
\end{figure}

The upper panel of Fig.~\ref{merTotDM} shows the evolution with redshift of 
the number density 
per unit volume
of events that produce starbursts, including major mergers, 
minor mergers with
starbursts, and disc instabilities.
As can be seen,
starbursts arising from major and minor mergers peak around $z\sim 2$ and $3$, respectively.
Disc
instability events increase in number as the redshift decreases,
with a number density that peaks at $z\sim 1$.  At lower redshifts, the number
of disc instabilities is higher than that of minor mergers.
The lower panels of Fig.~\ref{merTotDM} show the number of events 
that trigger starbursts
weighted by
the ratio between the mass of stars produced during the burst and
the stellar mass, for galaxies of high stellar mass, 
$M_{\star}>10^{10} M_{\odot}$ (lower left panel),
and low stellar mass, $M_{\star}<10^{10} M_{\odot}$ 
(lower right panel). 
It can be seen that, regardless of the stellar mass, 
disc instabilities are the preferred mechanism for the growth of stellar 
mass at all redshifts.  
Major and minor mergers play a minor role in the growth history 
of the stellar mass. 
An important point to notice is that the growth of stellar mass in high mass
systems reaches a peak at redshifts $z\sim 1.5$, whereas low stellar mass galaxies
continue to grow almost down to $z=0$.

\subsection{Black Hole growth, AGN outflows and gas cooling suppression}\label{BG_AGN}

As it was discussed before, there is growing evidence that AGN are extremely
important for galaxy formation and evolution. AGN are candidates to suppress
or reduce
cooling flows, thus affecting the star formation process and
associated galaxy properties, such as
the luminosity function,
colours, and stellar masses.
As it was first pointed out by \citet{Hoyle63}, 
the energy source of a quasar or AGN is gravitational and could arise
from a highly collapsed object or a massive black hole, an idea
broadly agreed to by most authors in the field.
Thus, in order to produce AGN, we first need to follow
the growth of central black holes within galaxies, for which we
propose a physical model based
on a combination of prescriptions given by \citet{Croton06}
and \citet{Malbon07}.
Considering that QSOs and AGN are different manifestations
of the same phenomenon,
the models of BH growth used in these works are motivated by
the observational evidence linking the global
star formation rate and the
optical QSO density in the universe (e.g. \citealt{boyle98}) in which both are
correlated.  
Given that there are two distinct ways to form stars, either
quiescently or in bursts, we also make a difference between the way in
which a BH grows in these two modes. We describe these 
modes 
separately in the
following two subsections.

\subsubsection{BH growth in violent star formation processes}

We assume that a BH grows from
perturbations to the gaseous disc resulting from galaxy mergers or disc 
instabilities.  
In such cases, cold gas is
funneled to the centre of the galaxy producing starbursts \citep{Croton06}.
The growth of BHs as a result of these processes is referred to as `starburst
mode growth'.
During a galaxy merger, we assume 
that the central BHs merge instantaneously.
For simplicity, we ignore the effects of
gravitational waves in BH mergers so that the resulting mass is the direct sum
of the two merging masses, 
with the BH mass in the satellite galaxy  
being transferred to the BH mass in the central one.
The resulting object continues to accrete cold gas. 
The mass accreted by the BH is proportional to the
total cold gas mass present; the efficiency of accretion is lower 
for unequal mergers. This is summarised by
\begin{equation}
        \Delta M_{\rm BH,mer}= f_{\rm BH} \frac{M^{\rm sat}}{M^{\rm central}}
\times
\frac{M_{\rm ColdGas}}{1+(200 \,{\rm km}\, {\rm s}^{-1}/V_{\rm vir})^{2}},
\label{massqsomode}
\end{equation}
where 
$M_{\rm ColdGas}$ 
is the cold gas mass
of the central galaxy after adding the cold gas of the satellite.
Note that we use a velocity of $200\,{\rm  km}\,{\rm s}^{-1}$ 
in the above formula instead of
a value of $280\,{\rm  km}\,{\rm s}^{-1}$ as considered by \citet{Croton06}.
As in that work, 
the value of the parameter $f_{\rm BH}$ is chosen so as to reproduce 
the observed local 
relation between the BH mass and the host bulge mass
($M_{\rm BH}-M_{\rm bulge}$ relation), as we 
discuss in Subsection \ref{BHBulge}; we adopt $f_{\rm BH}=0.015$.

The BH growth through disc instabilities is also represented
by Eq.~\ref{massqsomode},
where the ratio $M^{\rm sat}/M^{\rm central}$ is replaced by unity,
since this process is a dynamical self-interaction that depends on
the properties 
of a single galaxy. The behaviour of the BH growth during starbursts is shown in
Fig.~\ref{BHgrow},
and will be discussed in Subsection~\ref{BHGrowth}. 

It is important to note that there are no BH seeds in our model and,
therefore, the birth of central super massive BHs is triggered
when galaxies
undergo their first starbursts at some point in their evolution.

\subsubsection{BH growth in gas cooling processes} \label{BHgcoolsec}

The second way in which BHs are assumed to grow is from 
cold
gas accretion 
during gas cooling.  This occurs once a static hot gas halo has formed around 
the host galaxy. Note that satellite galaxies do not experiment this growth
mode since the reservoir of hot gas is associated only to the central galaxy.

Following \citet{Croton06}, we assume this gas accretion onto the BH to be
continuous and quiescent and to be described by a simple phenomenological model,
\begin{equation}
        \dot{M}_{\rm BH}= \kappa_{\rm AGN} \frac{M_{\rm BH}}{10^{8}
M_{\odot}} \times \frac{f_{\rm hot}}{0.1} \times \left(\frac{V_{\rm vir}}{200\, {\rm km}\,
{\rm s}^{-1}}\right)^{3} \label{massradio}.
\end{equation}
\noindent In this equation, $\dot{M}_{\rm BH}$ represents the black hole
accretion rate, $M_{\rm BH}$ is the black hole mass, $f_{\rm hot}$ is the
fraction of the total halo mass in the form of hot gas, $f_{\rm hot}=m_{\rm
HotGas}/M_{\rm vir}$, where $M_{\rm vir}$ is the virial mass of the host halo, and
$\kappa_{\rm AGN}$ is a free parameter. \citet{Croton06} note that $f_{\rm hot}$ 
becomes approximately constant when $V_{\rm
vir} \geq 150 \, {\rm km}\, {\rm s}^{-1}$; the dependence of $\dot{M}_{\rm
BH}$ on this quantity is not important. 

The mechanical heating generated by the BH
accretion (black hole luminosity) can be expressed, following the argument of
\citet{Soltan82}, as
\begin{equation}
L_{\rm BH}=\eta \, \dot{M}_{\rm BH} c^{2} \label{LBH},
\end{equation}
where $c$ is the speed of light, and $\eta$ is the standard 
efficiency of energy production that occurs
in the vicinity of the event horizon.  
This process may inject sufficient energy 
into the surrounding medium
to regulate gas cooling since it reduces, or
even stops, the cooling flow.
Consequently,
the cooling rate has to be
modified 
accordingly,
\begin{equation}
\dot{M}_{\rm cool}^{'}= \dot{M}_{\rm cool}-\frac{L_{\rm BH}}{V_{\rm vir}^{2}/2}
.
\label{coolAGN}
\end{equation}
This feedback process is referred to as `radio mode feedback',
following \citet{Croton06}; the way in which BHs grow during
this process is thus called  
`radio mode growth'. 

The BH luminosity depends on the combination of the free parameters $\kappa_{\rm
AGN}$ and
$\eta$ (Eq.~\ref{massradio} and \ref{LBH}, respectively). The local $M_{\rm
BH}-M_{\rm bulge}$ relation does not help in tuning these values since
it is not sensitive to them. However, the QSO and galaxy LFs are good
observational constraints.
We find that by adopting $\kappa_{\rm AGN} =
2.5\times10^{-4} \,{\rm M}_{\odot}\, {\rm yr}^{-1}$ and $\eta=0.1$, the observed
shape of the QSO LF and the bright-end break of the galaxy LF
 are fairly well reproduced as it is described in
Subsections~\ref{secQSOLF} and \ref{LumCol}, respectively.

Notice that we are not taking into account the fact that
BHs may be radiatively inefficient
when accreting gas during the
cooling flow process.
This possibility is supported by the presence of BHs at the centre of
clusters with cool cores that are
usually weak radio sources (e.g. M87 in the Virgo Cluster, 
\citealt{Le07},
\citealt{Wang08}).
While this particular phenomenon might
be accounted for by a non-constant efficiency $\eta$,
some models use lower BH accretion rates
instead
(e.g.
\citealt{Jolley07}, \citealt{Stern08}).

\subsubsection{Remarks on BH luminosity, $L_{\rm BH}$}

The definition of the black hole luminosity, $L_{\rm BH}$,
given in the previous section (Eq.~\ref{LBH}), 
involves the mass accretion rate onto the black hole.
This quantity is given by Eq.~\ref{massradio}
for the radio mode BH growth. In the case of 
the starburst mode growth, which occurs during mergers and disc 
instabilities, the mass accretion rate is estimated from 
the accreted BH mass in the starburst mode, as it is given
by Eq.~\ref{massqsomode}, 
 \begin{equation}
        \dot{M}_{\rm BH,SB}=\frac{\Delta M_{\rm BH,accr}}{\Delta T},
\label{massrateqso}
\end{equation}
where $\Delta T$ is the time interval determined by the
$50$ subdivisions between 
consecutive simulation snapshots, introduced to increase the accuracy of the integration of differential
equations;
this is the same time interval applied to estimate
$\Delta M_{\star}$ from Eq.~\ref{sfr}.

In this way, the luminosity of the black hole will grow every time the galaxy
undergoes star formation in mergers and disc instabilities, and during gas
cooling processes with soft, quiescent formation. 

\subsection{The Eddington Limit in the `BH' Luminosity}\label{Eddington}

For a black hole of mass $M_{\rm BH}$, the maximum allowed luminosity is
the Eddington luminosity, defined as
\begin{equation}
L_{\rm Edd}=4\pi G M_{\rm BH} \frac{c m_{\rm p}}{\sigma_{\rm T}} \label{Ledd},
\end{equation}
where G is the gravitational constant, 
$m_{\rm p}$ is the proton
mass and $\sigma_{\rm T}$ is the Thompson cross-section for the electron; 
the latter 
assumes a gas composed mostly by hydrogen.
In this expression, we consider the black hole mass 
at the end of the accretion process.
When the calculated black hole luminosity is greater than the corresponding
$L_{\rm Edd}$, we set $L_{\rm BH}=L_{\rm Edd}$. Nonetheless,
we do not limit the BH accreted mass, 
a choice justified by the lack of strong observational
constraints on the
efficiency $\eta$, that regulates the amount of accreted mass
that is transformed into black hole luminosity.
In our model, the maximum instantaneous fraction of BHs with
luminosities over the Eddington limit is only $\sim 6$\% at very high redshift,
so this assumption implies a minor effect. 

\section{Behaviour of the new model SAG2}\label{resultsec}

The principal aim of introducing AGN feedback in models of galaxy formation
is to account for the observed influence of AGN
on the evolution of galaxies.
In particular, the inclusion of AGN in 
semi-analytic models 
(\citealt{Cattaneo06}; \citealt{Menci06}; \citealt{Croton06}; \citealt{Bower06}; \citealt{Malbon07}; \citealt{Marulli08}) has
also proven to be a good way to avoid problems present in
previous versions of semi-analytic models such as an excess of bright, 
 extremely blue galaxies, when
compared to observations (as occurs in SAG1).

In the original model SAG1, a number of free parameters
was set to regulate the gas cooling, star formation, supernovae
feedback and galaxy mergers, and to
determine the circulation of metals
among the different baryonic components.
These parameters were carefully tuned in \citet{Cora06}
to satisfy numerous observational
constraints such as Milky Way properties, the luminosity
function, and the Tully-Fisher (TF), colour-magnitude and mass-metallicity 
relations.
An important success of the original code resides in that 
the implemented chemical model was able to explain the
spatial distribution of iron in
the intracluster medium 
at $z=0$,
and 
the temporal evolution of its mean iron content \citep{Cora08}.

Once the new free parameters involved in the black hole growth and AGN activity
are added, a new calibration of the model is performed to 
reproduce the above mentioned 
observational constraints
 plus new observational relationships
between the properties of black holes and their host galaxies, 
that is, BH mass-bulge mass, BH mass-bulge luminosity relations, 
and the QSO luminosity function.

In the remainder of this section, we 
analyse
the model predictions on
BH relations
and how the heating of AGN affects 
galaxy properties.

\subsection{Black hole growth}
\label{BHGrowth}

\begin{figure*}
\begin{center}
\hspace{-3em}
\includegraphics[width=0.8\textwidth]{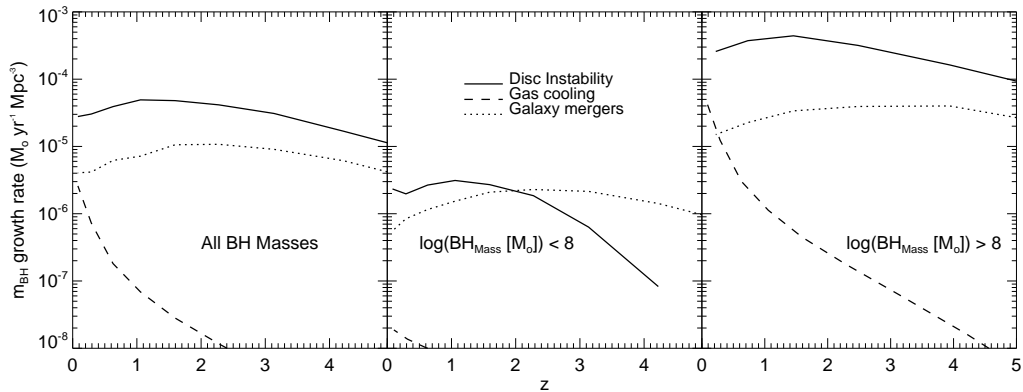}
\caption{Volume averaged individual BH growth rate as a
function of redshift for the three mechanisms that drive the BH growth: 
gas cooling
processes (dashed line), galaxy mergers (dotted line) and disc instabilities (solid line),
for all BHs masses (left panel), small black holes ($<10^{8} M_{\odot}$, 
middle panel) and massive black holes ($>10^{8} M_{\odot}$, right panel).}
\label{BHgrow}
\end{center}
\end{figure*}

The three different
processes driving the BH growth are gas cooling,
galaxy mergers and disc instabilities. Their influence on the volume
averaged individual BH growth is shown in Fig.~\ref{BHgrow}, 
represented by different
line styles.  The left panel involves all BH masses, showing that disc
instabilities are the major contributors to the BH growth at all redshifts, 
closely followed by
mergers, both with a smooth variation over a wide range of redshifts.
Gas cooling processes become more important at low redshifts with the 
growth rate steeply increasing by two orders of magnitude for
$z\la 2$.

These results differ from those reported by \citet{Bower06}, 
who find that mergers have a
small contribution to the BH growth, and that
disc instabilities are the most important contributors at high redshifts.  
\citet{Bower06}
also find that gas cooling becomes dominant at low redshifts, and that it operates since
earlier epochs.
On the other hand, \citet{Croton06} find that the radio mode growth is less important
than their `QSO mode' growth (associated with starbursts) in  contrast 
to the results from \citet{Bower06}, and that it stays constant since $z\sim 2$, a different behaviour
to that seen in our model. 
These differences demonstrate that the model implementation and parametrization
have a notorious influence on the detailed history of BH growth. In spite of
this, different semi-analytic models are still able to reproduce several
observational constraints involving BH masses, such as the `BH-bulge' relations 
described in the next Subsection\footnote{
 The different BH growth histories presented by these models will show important
effects in the global galaxy population at different redshifts.  
As we show in the remainder
of this paper, we use several galaxy statistics to help reduce 
the degeneracies between semi-analytic models.
}.  

Fig.~\ref{BHgrow} also shows the dependence of BH mass growth rate on the three
above-mentioned mechanisms for low ($M_{\rm BH}< 10^{8} M_{\odot}$, middle panel) and high
($M_{\rm BH}> 10^{8} M_{\odot}$, right panel) BH masses. 
For low mass black holes, we find that mergers dominate their growth down to $z \sim 2$,
where disc instabilities start to dominate for lower redshifts.
The effects of gas cooling processes are negligible. In the case of high BH masses, 
the behaviour of the three
mechanisms remains similar to what is shown by the full BH sample.

\begin{figure}
\begin {center}
\includegraphics[width=0.42\textwidth]{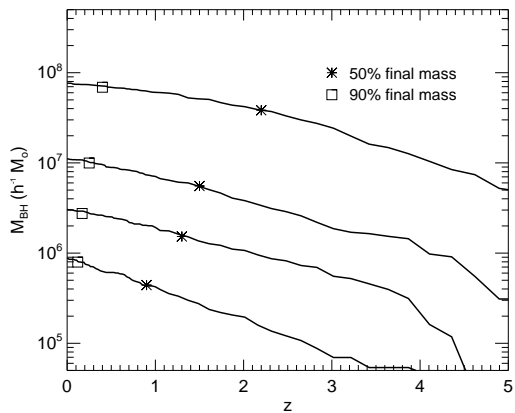}
\caption{Average growth history of BHs for different mass ranges starting 
from $z=5$. The tracks are constructed for 
a fixed number of BHs per mass bin ordered according to their average masses
(lines). 
The asterisks indicate the
points where the BHs have reached half of their final masses,
while empty squares represent the time 
when $90\%$ of the final mass has been accreted.}
\label{tracks}
\end {center}
\end{figure}

The average increase of BH mass with redshift for different ranges of BH
masses is shown in Fig.~\ref{tracks}. 
High mass BHs
acquire both, half and $90$ per cent of their final masses, at earlier times than 
smaller BHs. In other words, massive black holes form first. This is in
agreement with recent observational works (e.g.
\citealt{marconi04}) that find that, by $1.5\la z \la2.5$, massive black holes have
already attained at
least one half of their final mass; smaller BHs only reach half of their
final mass at $z\la1$. This is consistent with findings on
the evolution of the X-ray LF of QSOs (\citealt{cowie96}; \citealt{ueda03};
\citealt{Barger05}), where the number of bright
sources is higher at higher redshifts. Moreover, observations have 
shown that massive galaxies are already in
place in the distant universe (e.g. \citealt{Drory04}). With respect to star
forming galaxies, it has been found that their luminosities in the near 
{\em IR} bands decline
with decreasing redshift since $z \approx 1$ \citep{cowie96}. All these
observational evidences support the idea of an `antihierarchical' luminous
activity (star formation or mass accretion onto black holes) since it appears 
to occur earlier in more massive objects, 
a phenomenon generally termed `downsizing'.

The evolution of galaxies and BHs are closely connected, as indicated by the
BH-bulge relations (Section~\ref{BHBulge}). 
Therefore, the presence of downsizing in the growth of BHs
can be understood by taking into account
the fact that the star formation activity
in massive galaxies peaks at higher redshifts ($z \sim 1.7$)
than in less massive objects ($z \sim 0.5$, cf. Fig.~\ref{merTotDM}).
This phenomenon is reinforced 
by the effect of AGN feedback in masive systems that stops
the star formation activity earlier, and consequently the fueling of cold gas that feeds the BHs.
Downsizing has also been reported in 
previous semi-analytic models in the study of both
the growth of BH (\citealt{Bower06}, \citealt{Malbon07})
and of elliptical galaxies (\citealt{Cattaneo08}).

\subsection{`BH-bulge' relations}\label{BHBulge}

We require our model to reproduce as many observational constraints
on central BHs as possible.  We start by studying
the relations between the black hole mass and the host bulge properties:
the `BH-bulge' relations. 

\begin{figure}
\begin{center}
\includegraphics[width=0.43\textwidth]{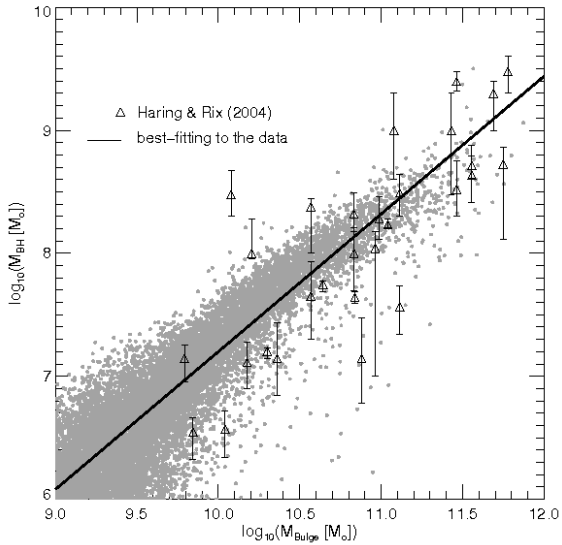}
\caption{Relation between the BH mass and the host bulge mass.
Model results are shown in grey dots. They are
compared to observations by
\citet{Haring04}, which are represented by triangles
with their corresponding error bars; the solid line
corresponds to the best-fitting power law to these data.}
\label{BH-B}
\end{center}
\end{figure}

The first comparison, and the most direct one, corresponds to the correlation
between BH mass and host bulge mass. In Fig.~\ref{BH-B}, we compare our model
results
(grey 
points) with observations by \citet{Haring04}, which are shown as triangles,
along with their best-fitting power law. In order
to obtain a good agreement with this observational relation, 
we vary the 
disc instability parameter, 
$\epsilon_{\rm disc}$, 
and the parameter $f_{\rm BH}$ (Eq.~\ref{eqDisc} and Eq.~\ref{massqsomode},
respectively). 
Even though black holes and bulges grow via 
the same processes,
a well adjusted correlation cannot be ensured
without a fine tuning of 
these parameters
since the BH growth not only depends
on the amount of available cold gas, as the bulge growth does,
but also 
on the BH mass and the virial velocity of the halo.

\begin{figure}
\begin{center}
\includegraphics[width=0.43\textwidth]{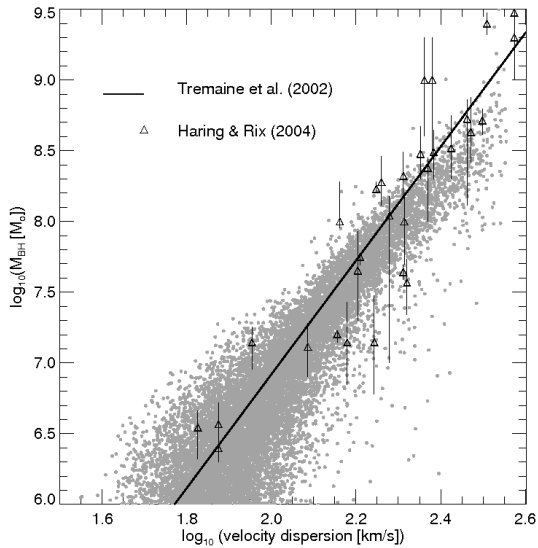}
\caption{Relation between the BH mass and 
host bulge velocity dispersion, $\sigma_{\rm Bulge}$. Model results are
represented by grey points. They are compared with observations of
\citet{Haring04} (triangles) and 
\citet{tremaine02} (solid line).}
\label{BH-sigma}
\end{center}
\end{figure}

\begin{figure}
\begin{center}
\includegraphics[width=0.43\textwidth]{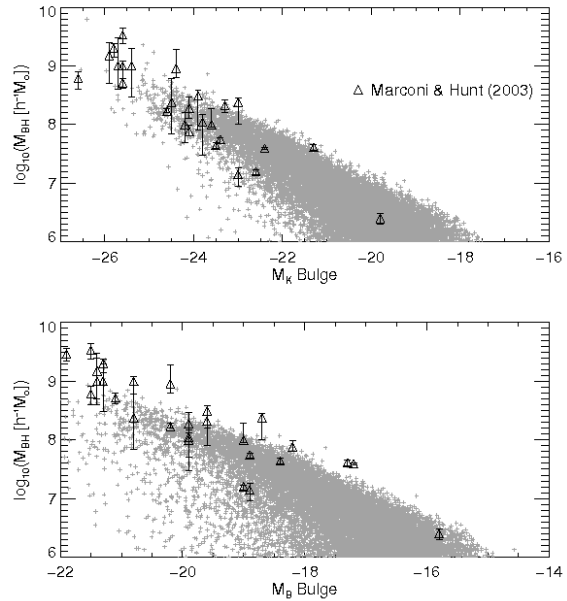}
\caption{Relation between the BH mass and bulge $B-$
and $K-$band absolute magnitudes.Model results are represented by grey points.
Compared with observations of
\citet{marconi03}
(black triangles).}
\label{BH-L}
\end{center}
\end{figure}

The relations between BH mass and bulge velocity dispersion, 
$\sigma_{\rm Bulge}$
(Fig.~\ref{BH-sigma}), and 
between BH mass and bulge luminosity 
(Fig.~\ref{BH-L})
also show excellent agreement with observations.
They are naturally obtained from the choice of model parameters motivated by the BH-Bulge mass relation.
This agreement is not always reached by other semi-analytic models;
for instance,
\citet{Malbon07} show a shallower 
relation between BH mass and bulge velocity dispersion
than the observed trend at 
high values of velocity dispersion, while our model produces a good match over a larger
dynamical range.

\subsection{QSO Luminosity Function}\label{secQSOLF}

Whenever a black hole accretes material there is an associated
bolometric luminosity and, therefore, an active galactic nucleus. 
Several clues on the nature of AGN are provided
by the study of the evolution of the QSO luminosity function.

In order to compare the QSO luminosity function 
obtained from our model with observational results, we estimate
a QSO absolute magnitude in the optical band $M_{\rm bJ}$ (blue magnitude in Johnson's system),
following Eq.~\ref{eqQbj} of \citet{Croom05},
\begin{equation}
   M_{\rm bJ}=-2.66 \,{\rm log}_{10}(L_{\rm Bolom}^{\rm QSO}) + 79.42,
    \label{eqQbj}
\end{equation}
where $L_{\rm Bolom}^{\rm QSO}$
is the bolometric luminosity in units of Watts
given by the black hole luminosity defined
by Eq.~\ref{LBH}.
We then build the QSO LF at different redshifts and compare
its evolution with 
measurements by \citet{Croom04} and \citet{wolf03}.
The former work presents results for the QSO LF from
the 2dF QSO redshift survey (2QZ)
over the redshift range $0.4 \la z \la 2.1$, while the latter
uses the COMBO-17 survey that
contains a large sample of faint QSOs in the redshift range $1.2 \la z
\la 4.8$.  
Note that the LF found
by 
\citet{wolf03}
is consistent with results 
obtained
from surveys such as SDSS and 2QZ.

\begin{figure}
\begin {center}
\includegraphics[width=0.45\textwidth]{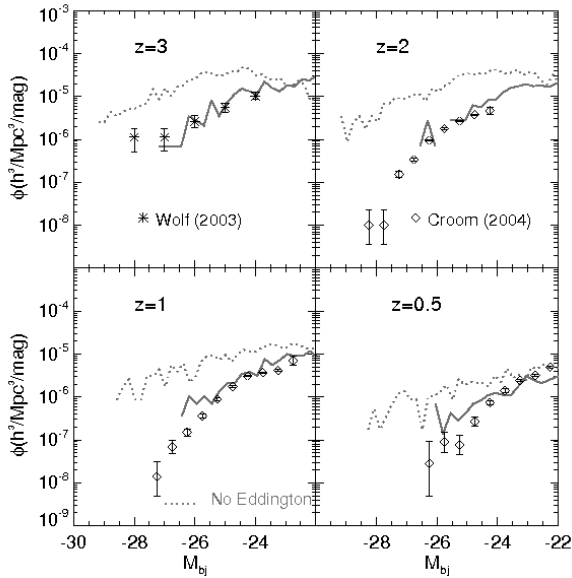}
\caption{Model QSO LF compared with observational data 
by \citet{Croom04} for $z\la 2$,
and \citet{wolf03} at $z\approx 3$. The grey solid lines
show the final model including Eddington limit with the parameters indicated 
in Section~\ref{Modelsec}. 
To obtain good
agreement with observations it was necessary to assume a fraction,
$f_{\rm obsc}$=0.8 of obscured QSOs. The resulting model QSO LF with
no Eddington Limit is represented by grey dotted lines.}
\label{LFQSO}
\end {center}
\end{figure}

The QSO luminosity function is shown in Fig.~\ref{LFQSO}.
The results of our model (solid lines) are in good agreement with the
observational data
of \citet{wolf03} and \citet{Croom04}, represented by symbols. 
These results are obtained by the inclusion of the Eddington limit
(Subsection \ref{Eddington}) and a fraction of obscured QSOs.  The application
of the Eddington limit
helps to recover the observed
QSO LF, although a constant normalization offset remains at all redshifts.  In order
to solve this discrepancy, we
find important to consider that a fraction $f_{\rm obsc}$ of model QSOs are obscured by the gas torus
assumed to be present in
the `unified model' of AGN galaxies (e.g. \citealt{fanaroff74}; 
\citealt{barthel89}; \citealt{madau94}; \citealt{gunn99});
we adopt $f_{\rm obsc}=0.8$.
This fraction is consistent with  
observational results 
(\citealt{gunn99}; \citealt{lacy07}) which show that a large number 
(possibly the majority) of QSOs
are obscured.

Model results with
$f_{\rm obsc}=0.8$ and
no Eddington Limit 
are represented as grey dotted lines,
showing that 
this limit produces changes up to a factor of $\sim 6$ in the QSO luminosity.
\citet{Malbon07} also 
find that the Eddington limit contributes to obtain
a good agreement with observations. We remark that our simulation 
box size is only mid-size and, therefore, does not allow to reproduce
the entire dynamic range of observed QSO number densities.

We stress the fact that we are using a fixed value of $f_{\rm obsc}$ at
all redshifts; therefore, it can be said that QSO luminosities in our
model depend exclusively on the physical properties of the AGN phenomenon. 
We notice, however, that at low redshifts the model shows slightly 
higher number densities than the data,
possibly implying a larger obscured 
fraction as the redshift decreases.  There is some observational evidence 
in this direction, where the fraction of obscured
QSOs becomes higher at $z\la1$ 
(e.g. \citealt{franceschini02}; \citealt{gandhi03}).
However, this redshift dependence is not always found
(e.g. \citealt{ueda03}; \citealt{Gilli03}), being necessary larger
samples to understand this trend.
On the other hand, the results obtained by \citet{ueda03} 
imply that the fraction of absorbed AGN decreases at high
luminosities.  Our model shows the opposite behaviour; 
a larger fraction of obscured AGN is required at the bright-end of the 
LF to further improve the agreement with the observational data.

As we have mentioned in Section~\ref{BHgcoolsec}, 
the shape of the QSO LF is sensitive to the parameter $\kappa_{\rm AGN}$,
responsible for the regulation of the gas accretion rate onto BHs via gas cooling
processes.  Adopting larger values of $\kappa_{\rm AGN}$ 
leads to a higher QSO number density. This increment occurs at high redshifts ($z\gtrsim1$)
for  low and intermediate luminosities, while at lower redshifts it 
mainly affects the bright-end of the QSO LF. 
The latter arises as a consequence of the increased impact
of gas cooling processes on the growth of massive
BHs at low redshifts.
It is important to remark
that we do not make assumptions about the QSO lifetime since we consider the
BH luminosity as the instantaneous rate of energy ejected from the BH.

\subsection{Galaxy properties: effects of AGN feedback}

Semi-analytic models are calibrated to reproduce, as best as possible,
properties of galaxies in the local Universe such as 
the luminosity function,
the TF relation, colour-magnitude diagram and gas fractions. 
The luminosity functions in different colour bands
obtained from SAG1 are characterized by an
excess of bright galaxies with respect to the observed 
luminosity functions.
The TF relation is found to lie within the observational error determinations,
following the best-fitting estimation by \citet{Giova97}.
The colour-magnitude relation for early type galaxies managed a
very good agreement with
the observational fitting by \citet{Bower92}. Gas fractions were also
found to be in very good
agreement with observational results by \citet{McGaugh97}.

In this section, 
we analyze the advantages of the new model SAG2 which now includes AGN feedback,
evaluating the influence of this process 
on different galaxy properties, and comparing 
with the results provided by the original version SAG1.

\subsubsection{Luminosity and stellar mass functions}\label{LumCol}

\begin{figure}
\begin {center}
\includegraphics[width=0.45\textwidth]{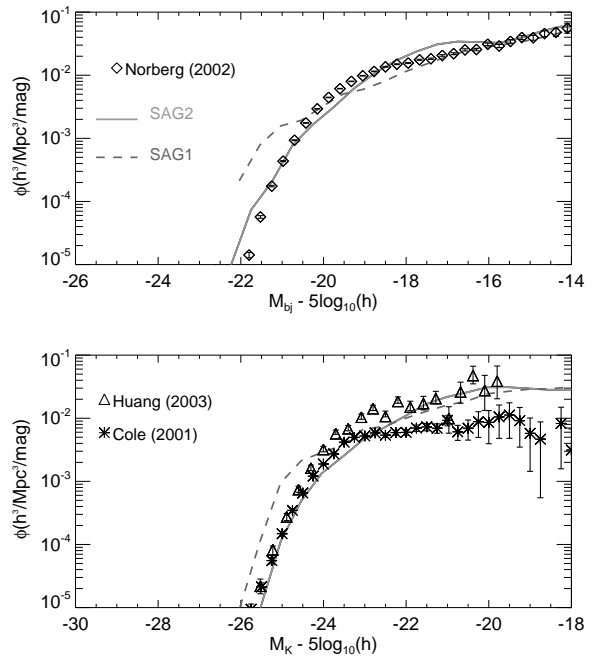}
\caption{{\em Upper panel:} $b_{\rm J}$ band galaxy LF.  
{\em Lower-panel}: $K$-band galaxy LF.
Solid lines represent results from our new model, SAG2, while dashed lines
correspond to the previous version, SAG1; different symbols
identify observational LF estimates by \citet{Norberg02},
\citet{huang03} and \citet{Cole01}.}
\label{galLF}
\end {center}
\end{figure}

Luminosity functions in the $b_{\rm J}$- and $K$-bands are shown for SAG2
galaxies in Fig.~\ref{galLF}. It
can be seen that in both cases the model closely follows the trend denoted by the
observational results from \citet{Norberg02} for the $b_{\rm J}$-band, and \citet{huang03}
and \citet{Cole01} for the $K$-band.  At the bright end of the LF, the agreement with observations 
has improved significantly over the previous version of the model, SAG1.
This improvement comes from a 
more vigorous quenching of cooling of hot gas at low
redshifts operating in SAG2, which prevents the presence of very luminous massive
galaxies. 

\begin{figure}
\begin {center}
\includegraphics[width=0.42\textwidth]{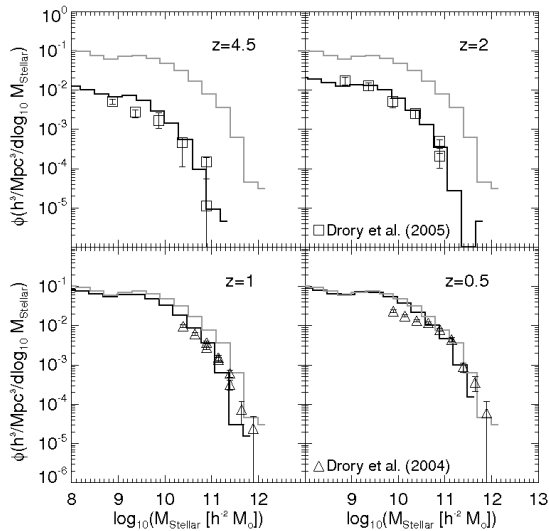}
\caption{SMF for the model galaxies in SAG2
at redshifts $z = 4.5$, $2$, $1$ and $0.5$ (black solid lines); for comparison, 
all panels show the z=0 model SMF (grey solid lines). Symbols represent
observational data from \citet{Drory04} and \citet{Drory05}.}
\label{SMF}
\end {center}
\end{figure}

The masses of model galaxies are able to reproduce the observed evolution of
the stellar mass function (SMF).
The dependence of the SMF with redshift is shown in
Fig.~\ref{SMF}. Here we compare the SMF from our model 
SAG2 with observations from \citet{Drory04}
and \citet{Drory05} in four redshift bins, from $z\simeq 0.5$ to $4.5$;
model results are represented by black solid lines. 
The model SMF is in good agreement with observations 
at all redshifts. 
The grey lines show the $z=0$ model SMF, and help to illustrate the lack of
significant evolution in the SMF from $z \sim 0.5$.

Note that our model uses a Salpeter IMF; certain models consider different IMFs 
(e.g. \citealt{Baugh04} use a top-heavy IMF) 
in order to reproduce the
observational results.
It is important to remark that \citet{Drory04} and \citet{Drory05} also assumed a
Salpeter IMF to apply photometric techniques to determine redshifts 
and stellar masses of the best-fitting 
templates of galaxy spectra. They also analyse
two independent fields, the FORS deep field and the GOODS-S field,
implying that the data
take into account cosmic variance to some degree and
are therefore representative of the observed universe at low and high redshifts.

\subsubsection{Galaxy morphology}
\label{ssec:morph}

We analyse  the morphology of SAG2 galaxies at $z=0$.
In order to determine a fraction of
morphological types from the model, we use the ratio between the bulge mass and the total
stellar mass, r=$M_{\rm Bulge}/M_{\star}$, 
as proposed by Bertone, De Lucia \& Thomas (2007) for this type of
analysis. We adopt a threshold between ellipticals and spirals $r_{\rm
thres}=0.95$,
classifying as ellipticals all bulge-dominated systems with 
$r>r_{\rm thres}$, and as spirals
those with $0<r<r_{\rm thres}$;
systems with no bulge (r=0) are classified as irregulars. 
The threshold between ellipticals and spirals is not firmly established;
for instance, \citet{bertone07} use $r_{\rm thres}=0.7$. 
Nonetheless, a higher
threshold, as the one adopted here, defines very well the 
elliptical population,
allowing them to have at the most a stellar disc of $5$ per cent 
the total mass.  

\begin{figure}
\begin {center}
\includegraphics[width=0.42\textwidth]{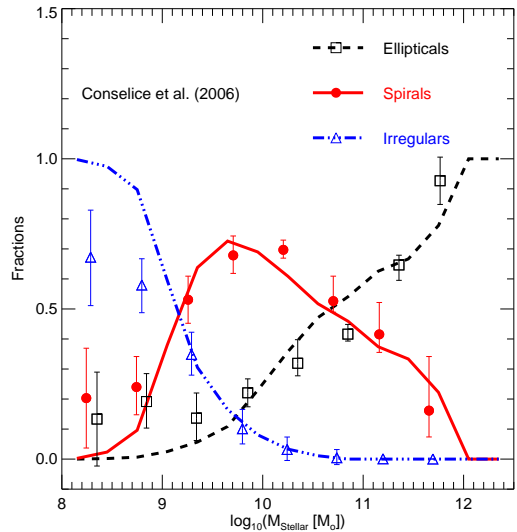}
\caption{Fractions of SAG2 galaxies of different morphological types 
at $z=0$ as a function of stellar masses.
Model results (lines) are compared with 
observations taken from \citet{Conselice06} (symbols):
irregular galaxies are represented by  
dashed-dotted lines and triangles, spirals 
by
solid lines and filled circles, and ellipticals 
by
dashed
lines and open squares.}
\label{morpho}
\end {center}
\end{figure}

Fig.~\ref{morpho} shows the fractions of morphological
types as a function of stellar mass 
for our model (represented by lines) and the
observational data
by \citet{Conselice06}, represented by symbols with errorbars. The data include
a sample of more than $22,000$
galaxies at $z\la 0.05$ with a visual morphological classification
from the RC3 catalogue \citep{vaucu91b}. 
We can appreciate that the fractions of different morphological types in the
model agree very well with the observations over the whole stellar mass range.
SAG2 shows a clear improvement over the previous version of our model, SAG1, in which
$\sim 80$ per cent of massive galaxies evolved to be spirals.
Starbursts triggered by disc instabilities are a key ingredient to achieve a good
agreement with the observed morphological types; disc structures are destroyed during these events
thus favouring the formation of early type galaxies. When including disc
instabilities in SAG1 the percentage of high stellar mass spirals is reduced to $\sim 40$ per
cent.
 
Changes in $r_{\rm thres}$, which we varied from
$0.7$ and $0.95$,
produce a shift in the height of the spiral fraction peak from $\simeq 0.55$ to
$\simeq 0.75$.  Spiral galaxies at intermediate stellar masses are
the more sensitive population to $r_{\rm thres}$. Elliptical galaxies are the dominant type at
high stellar masses regardless of the value of this parameter.

In order to test the reliability of the morphological classification  
described above, we 
apply the criteria given
by \citet{DeLucia04}, in which spiral galaxies satisfy the condition 
$1.2 <M_{\rm B, \rm Bulge}-M_{\rm B, \rm Gal}< 2.4$;
we also consider
the definition of Hubble types given by \citet{Springel01}.
Our conclusions are not significantly affected in any of these cases.

Recent works on the analysis of morphological types in semi-analytic galaxies 
obtain slightly different results.
\citet{DeLucia07}  fail to predict the presence of high mass
spiral galaxies.
On the other hand, \citet{bertone07} show a good agreement with observations 
at all stellar masses although their model produces
a significant excess of bright galaxies in the LF and a smooth colour distribution with
most of their galaxies residing in a red sequence.
The morphological aspect of model galaxies is quite sensitive to
the different parametrizations of the processes
affecting the star formation, 
including the effects of disc instabilities
and major and minor mergers. Thus, it is not surprising
that different semi-analytic codes lead to
different dependencies of morphological type fractions with stellar mass.

\subsubsection{Galaxy colours}

\begin{figure*}
\begin {center}
\includegraphics[width=0.33\textwidth]{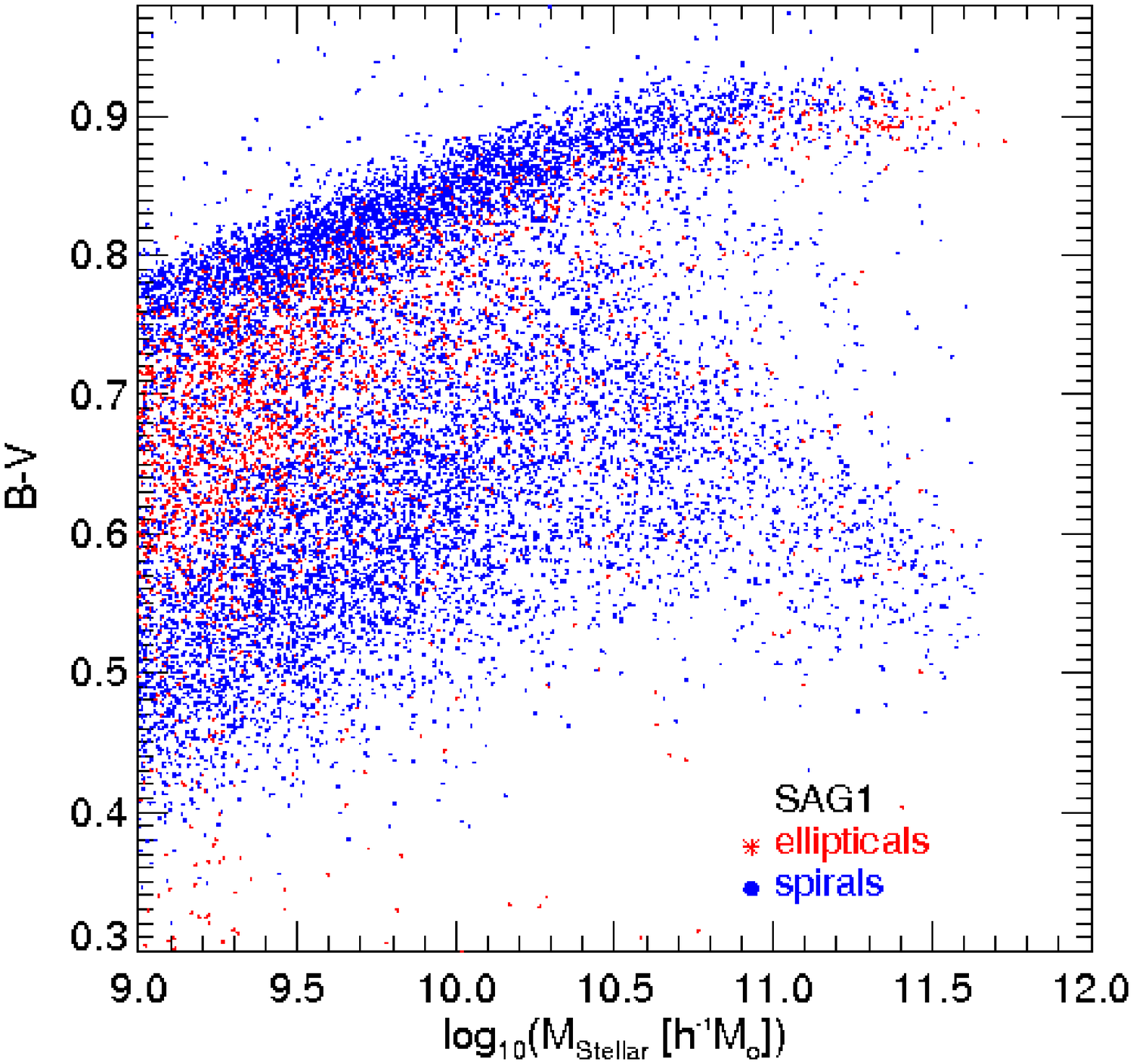}
\includegraphics[width=0.33\textwidth]{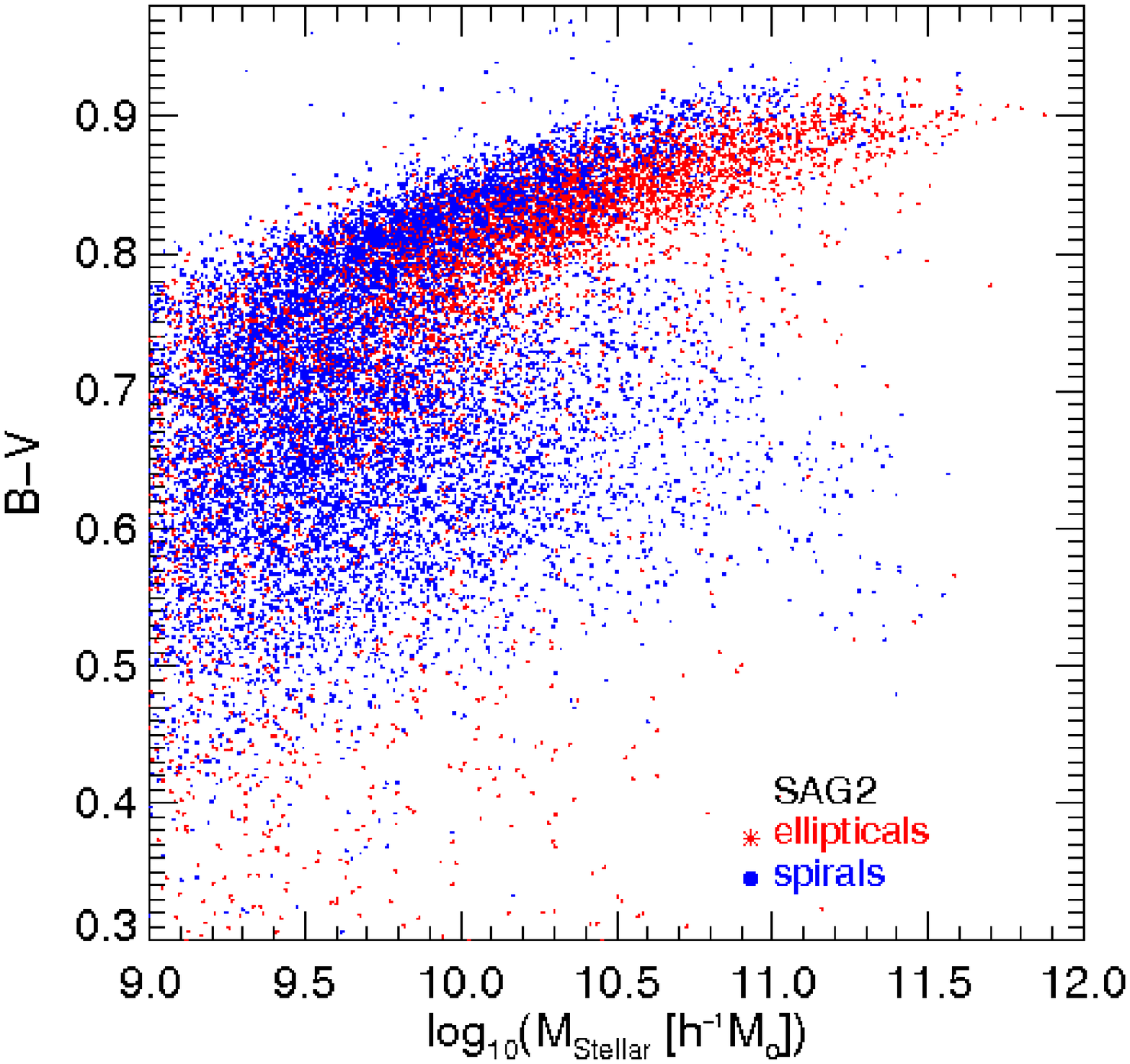}
\includegraphics[width=0.33\textwidth]{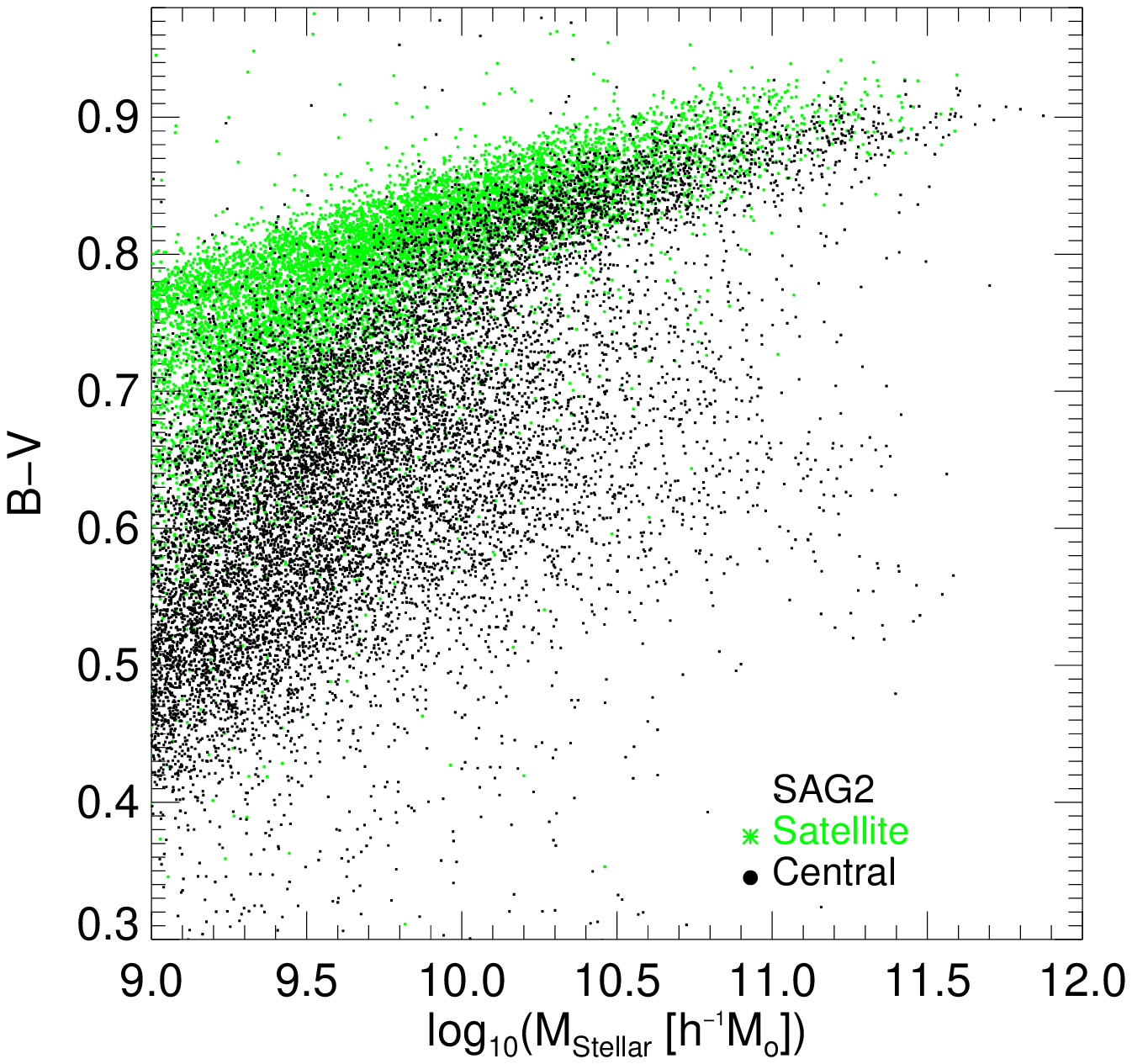}
\caption{Colour distribution for ellipticals (red symbols) and spirals
(blue symbols) type galaxies obtained from models SAG1
(left panel), and SAG2 (middle panel).  SAG2 galaxies separated into 
centrals and satellites are also shown (right panel, black and green symbols,
respectively).}
\label{colour-distribution}
\end {center}
\end{figure*}

The AGN feedback has an important effect on galaxy colours.
Fig.~\ref{colour-distribution} presents $B-V$ colours as a function of 
stellar mass for the models SAG1 (left panel) and SAG2 (middle panel),
where galaxies are separated in ellipticals (red symbols) and 
spirals (blue symbols) following the criteria outlined in Subsection 
\ref{ssec:morph}.  The right panel shows the same
SAG2 galaxy population splitted into central (black symbols) and satellite galaxies (green
symbols).

SAG2 galaxies show that the inclusion of AGN feedback produces
galaxies with $B-V$ colours qualitatively compatible with measured
colour distributions (see for instance \citealt{Baldry06}) including
a bimodal behaviour and a negligible population of massive ($\gtrsim 10^{11}\, 
h^{-1}\,M_{\odot}$) blue galaxies.
This is a clear difference with respect to SAG1 where there is a population
of very massive, blue spiral galaxies.  As it was mentioned in Subsection 
\ref{ssec:morph}, this is a consequence 
of the generation of new spheroids by the action of the disc instability process
included in SAG2. 

The right panel of Fig.~\ref{colour-distribution} indicates that the red
branch is mainly composed by satellite galaxies, characterized by the fact that
their hot gas reservoir has been stripped when they were accreted by the central halo, 
aging quiescently until merging with a central galaxy.
Massive central galaxies also become red as a result of
a combination of different effects.
On the one hand, the reduction or suppression 
of gas cooling leaves a galaxy with a limited cold gas
reservoir that is eventually exhausted; it then
undergoes a passive evolution unless a merger
with another galaxy takes place.  However, the possibility that this occurs
decreases at low redshifts, as it is shown in Fig.~\ref{merTotDM}.
On the other hand, if a merger occurs, it is very unlikely that
satellite galaxies will contribute enough cold gas to produce
a starburst that will appreciably change the colour and mass content 
of the central galaxy, since the gas cooling process does not occur in satellites.
Hence, mergers would more likely only increase the mass of 
the stellar population without giving place to an important
generation of new stars.

\section{Further improvements to the semi-analytic model of galaxy
formation}\label{Furthersec}

In this section, we explore the effects of two new ingredients that can be
added to the model.  We first take into account an additional 
form of feedback from AGN, associated with the black hole
growth during starbursts.  Then, we focus on the analysis of
the effects of different prescriptions for the dynamical friction time-scales 
affecting galaxy mergers.

\subsection{AGN Feedback in starburst events}\label{FeedQSOsec}

The importance of AGN feedback in galaxy evolution is undeniable, 
as theoretical and observational works indicate, 
although this is still a poorly understood phenomenon and 
is at present a matter of debate. It is mostly
accepted that the quiescent
gas accretion produces feedback from the active
nucleus, but there is no general consensus on the fate of the energy produced by
the BH mass accretion during starbursts.
Therefore, in this section, we use our new semi-analytic model 
SAG2 to explore 
the effects of AGN feedback during violent events (i.e., starbursts)
on galaxy properties. 

There are observational clues about the influence of AGN feedback in
starbursts. 
\citet{Donahue05} propose a dichotomy in galaxy clusters between those which are
radio active and show strong 
central
temperature gradients (indicating the action of
AGN feedback), and radio quiet 
with no temperature gradients.
Nonetheless, \citet{gastadello07} find a
counter-example in the cluster AWM 4 characterized by the absence of a
temperature gradient, as indicated by the XMM data, and the presence of an
active radio source at 1.4 GHz. The combination of these features cannot be explained if 
AGN activity is regulated only by feedback from gas cooling processes. 

We explore the effects of AGN feedback during starbursts
following two different possibilities: (i) the outflowing material 
is directly transferred to the hot gas surrounding galaxy, and (ii) 
the energy produced is high enough to expel material out
from the galaxy halo without affecting any baryonic component
associated to the galaxy.
We refer to these two mechanisms as retention and ejection modes, respectively;
the former has been used in both SNe feedback and AGN feedback
in radio mode throughout this work. 
The absence of definitive evidence favoring one scenario or the other only
allows us to analyse and compare the outcomes from each option.
Note that the ejection mode does not operate exactly in the same way as in,
for example, \citet{DeLucia04}, where the ejected gas is eventually
reincorporated to the dark halo.
AGN feedback in starburst mode 
heats the cold gas already present in the galaxy with a rate
$2 \times L_{\rm BH} / V_{\rm Vir}^{2}$.
The black hole luminosity, $L_{\rm BH}$, given by Eq.~\ref{coolAGN},
now involves the mass accreted in the starburst mode, as given by
Eq.~\ref{massrateqso}.

\begin{figure}
\begin {center}
\includegraphics[width=0.43\textwidth]{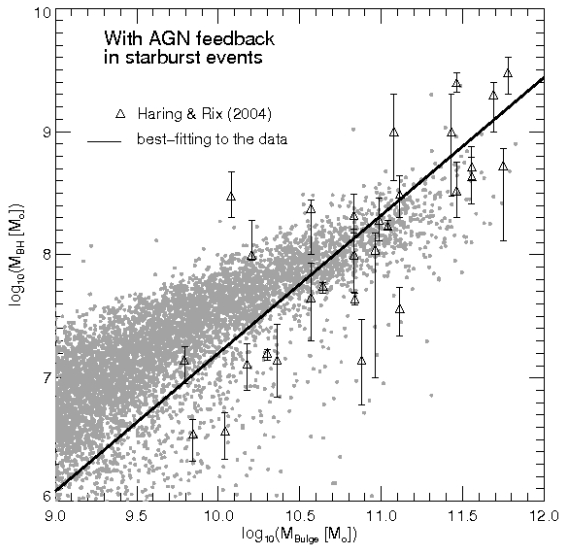}
\caption{Relation between BH and bulge masses for 
model results including AGN feedback during starbursts
in the 
ejection mode  
(grey points)
compared with 
observational estimates by \citet{Haring04} (triangles).}
\label{Bh-bulgeBothModes}
\end {center}
\end{figure}

Fig.~\ref{Bh-bulgeBothModes} shows the relation between black hole and bulge
masses considering AGN feedback produced when the
accretion occurs as a result of starburst events, for the case of the ejection
mode. The results from the retention mode are
not shown in this figure since the general trend does not vary
appreciably with respect to the ejection scenario,
and only produces a larger scatter of BH masses.
As can be seen, AGN feedback during
starbursts produces only slight changes with respect to the model where only
radio mode AGN feedback is applied (see Fig.~\ref{BH-B}).  
The main difference is a shallower relation produced by the complementary
effect of AGN feedback in starbursts, that reduces even more the available cold gas for bulge
growth so that, for small BHs, bulge masses are smaller.  Note
that this does not affect BH masses as significantly since BHs grow mainly through mergers and
disc instabilities (cf. Fig.  ~\ref{BHgrow}).
In any case, the
poor statistics in the observational relation does not allow us to determine which
of the three versions of our model (AGN feedback from gas cooling, or
AGN feedback during starbursts for the retention/ejection mode)
provides the best agreement.

\begin{figure}
\begin {center}
\includegraphics[width=0.5\textwidth]{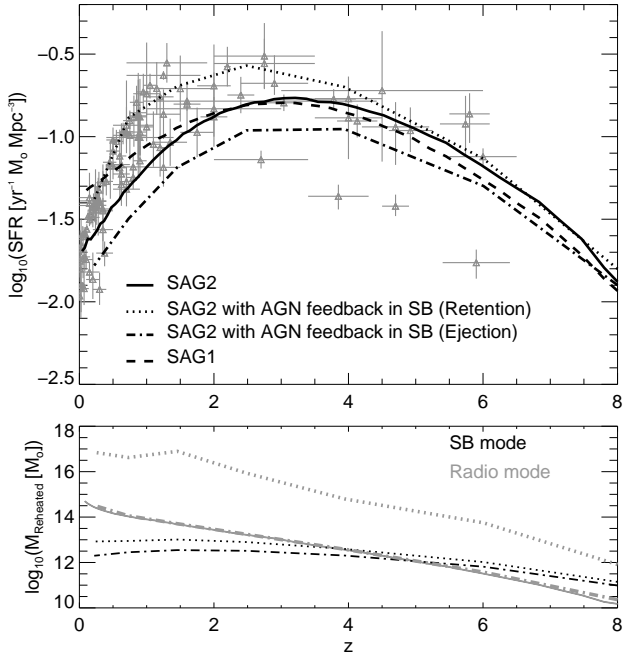}
\caption{{\em Upper panel}: Cosmic star formation rate for 
SAG2 (solid line), SAG2 including AGN feedback during starburst (with the same
parametrization as SAG2; 
dotted and dot-dashed lines for the retention and 
ejection modes, respectively) and SAG1 (dashed line), compared to observations compiled 
by \citet{hopkins06} (grey triangles with errorbars). 
{\em Lower panel}: Reheated mass from the different feedback mechanisms as a
function of redshift for the different AGN feedback models.
The grey lines represent the radio mode and the black lines the starburst mode.}
\label{SFRobserved}
\end {center}
\end{figure}

We have to take into account that 
AGN feedback in starburst mode is an additional ingredient that changes 
the mass and metallicity of the hot gas, 
which in turn
affects the cooling rate and thus the stellar formation rates. 
Fig.~\ref{SFRobserved} shows the global cosmic star formation
rate as a function of redshift for the models SAG1 (dashed line), SAG2 (solid
line) and SAG2 including AGN feedback in starbursts for the retention (dotted line)
and ejection (dot-dashed line) modes.  The latter two models use the same
set of parameters as SAG2.  Comparing
SAG2 with SAG1, we can appreciate that, for $z \gtrsim 3$, the SFR is larger in SAG2 
as a result of the disc instability and starbursts in minor
mergers included in this model. The feedback in radio mode becomes progressively
more effective for lower redshifts, increasing the reheated mass (solid grey line in
lower panel of Fig.~\ref{SFRobserved}) and appreciably
reducing the SFR.  When including AGN feedback in starbursts in the ejection mode, the SFR is
reduced at all redshift both with respect to SAG1 and SAG2, keeping the similar
redshift dependence as in SAG2. This is a consequence of the removal of
gas by the ejection mode, which is no longer available to cool and form stars.
In the case of the retention mode for the AGN feedback in starbursts,
the gas is mainly reheated by the
radio mode, as shown in the lower panel of Fig.~\ref{SFRobserved}. This is due to
a larger reservoir of hot gas available for gas cooling due to the retention of the
gas heated by the AGN.  This produces a much higher SFR in this mode with respect 
to the other models shown in the figure.

Three models, SAG1, SAG2, and SAG2 with retention mode AGN feedback in starbursts, 
show a reasonable agreement with the observational data compiled by
\citet{hopkins06}
(grey symbols with error bars) at all redshifts. However, SAG1 results show
a too high SFR at $z=0$.  
On the other hand, SAG2 with AGN feedback in starbursts in the ejection mode, shows
a much lower star formation history (SFH) at all redshifts, although it should be noticed that some observational
data points are consistent with even lower values of SFR at $z\gtrsim 2$.
Note that SAG2 provides a good agreement with the oberved SFH at all redshifts.  Only
at $z\simeq 1$ the model shows slightly lower values than the distribution
of observational results. However, recent studies by \citet{Wilkins08} show that
direct measurements of star formation rates 
(e.g. \citealt{hopkins06}) seem to be overestimated with respect to the SFR values
obtained from stellar mass functions.
Taking this into account, the agreement between our model SAG2 and observations 
at $z \simeq 1$ would be even more consistent.

\begin{figure}
\begin {center}
\includegraphics[width=0.45\textwidth]{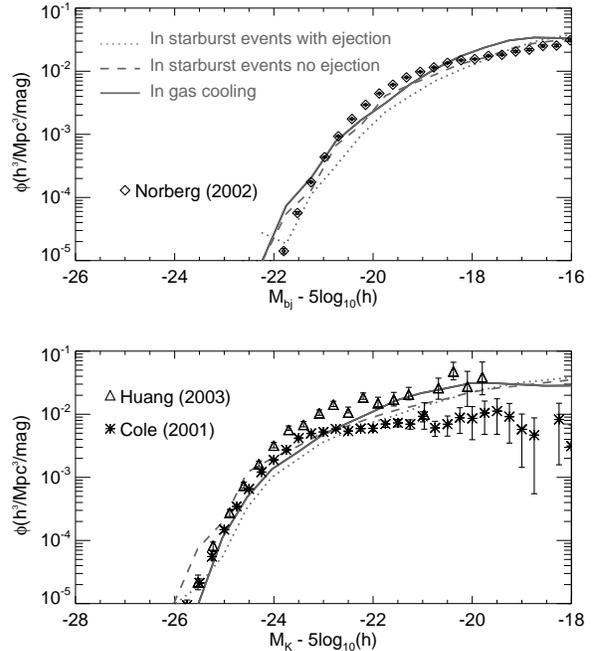}
\caption{$b_{J}$- (upper panel) and $K$-band (lower panel) luminosity functions
for three models: when AGN feedback is acting only during gas cooling
process (solid line), when AGN feedback is also produced in starburst 
events with ejection of the reheated material (dotted line); 
and when the material is
retained in the dark halo in which the galaxy resides (dashed line). 
 In the ejection mode we use the same set of parameters as in SAG2, while 
a value of $\kappa_{\rm AGN}= 2 \times 10^{-3} \,{\rm
M}_{\odot}\,
{\rm yr}^{-1}$ is adopted in the retention mode.}
\label{LFcompModes}
\end {center}
\end{figure}

We also analyse how the inclusion of AGN feedback in starburst mode
 affects the galaxy 
luminosity function. Fig.~\ref{LFcompModes} shows the $b_{\rm J}$- (top panel) and $K$-band 
(bottom panel) LFs for the three different models:
SAG2 (solid line), and 
SAG2 including feedback in starbursts in the retention (dashed line) and
ejection modes (dotted line).  As can be seen in Fig.~\ref{SFRobserved}, 
SAG2 gives lower SFRs when AGN feedback in starbursts with ejection is included. 
Therefore, in this case the LF is slightly shifted towards fainter luminosities by up
to half a magnitude. As the agreement with observed results is not badly affected by this,
we do not need to recalibrate the model parameters in this case.
In the retention mode,
the radio mode feedback efficiency, the $\kappa_{\rm AGN}$ parameter (Eq.~\ref{massradio}), needs to be 
increased in order to reproduce
the observed bright-end of the LFs; we adopt $\kappa_{\rm AGN}= 2 \times 10^{-3} \,{\rm
M}_{\odot}\,
{\rm yr}^{-1}$. After applying this
recalibration, the retention mode 
produces a good agreement with observations.
 
The morphology of model galaxies (estimated as in Section 
\ref{ssec:morph}) is affected by the AGN 
feedback during starbursts.
The ejection mode improves the agreement with the observed morphological fractions at low
stellar masses \citep{Conselice06}, producing a non-zero spiral population at
this mass range.  This mode also mantains the good agreement at intermediate and high stellar
masses shown by SAG2. On the other hand, the retention mode fails to reproduce the
observed morphological fractions at all masses, even after the feedback
efficency, $\kappa_{\rm AGN}$, is increased in order to match the observed galaxy
LF.      

The failure of the 
retention mode can be used as an indication that, if present, 
AGN feedback during starbursts should be strong enough to
expell the reheated material away from the galaxy halo. 
The only drawback to this ejection model is that it may
produce slightly low values for the SFH 
with respect to the observational trend, 
which in any case still shows data points that lie below the model
prediction at high, $z>2$, redshifts.

The implementation of AGN feedback in starbursts is
still crude since it considers the same efficiency as in the AGN feedback acting during 
gas cooling.  It is possible though,
that the physical considerations 
needed for these two different accretion processes, via gas cooling and during
starbursts, are not the same. 
For instance, 
instead of using different BHs feeding mechanisms to distinguish between
different feedback modes, the model described by \citet{Sijacki07}
uses a threshold value for BH accretion rate
above of which the feedback is similar to our starburst mode although with a
lower efficiency. We leave the detailed study of AGN feedback during starbursts 
to a forthcoming paper.

\subsection{New prescriptions for merger time-scales}\label{NMsec}

All the results presented in this work have been obtained considering
that galaxy mergers are driven by the dynamical friction process,
for which we have adopted 
the prescription given by Lacey \& Cole (1993, hereafter LC93).
We now take into account the new dynamical friction time-scales
proposed by \citet{jiang07} and \citet{boylan07} (hereafter J07 and B07, respectively).  
Both studies use numerical simulations to
measure merger time-scales and compare them to theoretical predictions based
on the Chandrasekhar formula. They find that, in contrast
to \citet{NFW97}, the theoretical merger time-scales are not in agreement with
the simulated values.  Equations (5) from J07 and (5) from B07 provide new 
fittings to the merger
time-scales that we test in our model.  The difference between these two 
fittings resides in
the weaker dependence on circularity in the prescription of J07 compared with
the one by B07. The overall conclusions from both studies
are the same, that the LC93 merger time-scales are underestimated for minor mergers and
overestimated for major mergers.
                                                                                
\begin{figure}
\begin {center}
\includegraphics[width=0.45\textwidth]{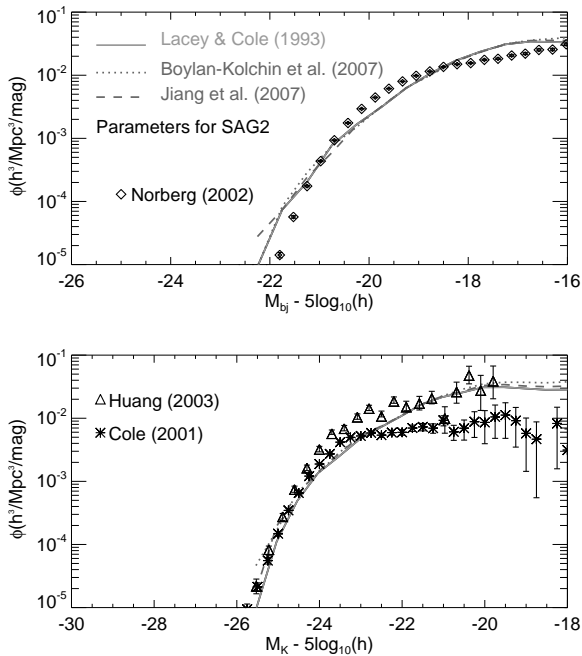}
\caption{$b_{\rm J}$- and $K$-bands galaxy luminosity function for three different
merger time-scales; the results using the prescription by \citet{lacey93} are
shown by solid lines,
those from \citet{jiang07} and \citet{boylan07} are represented
by dashed and dotted lines, respectively.}
\label{LFcomp}
\end {center}
\end{figure}

The effects of the three different merger time-scales on the galaxy population
can be readily seen in Fig.~\ref{LFcomp}, where we show the luminosity
function in the $b_{\rm J}$- and $K$-bands (upper and lower panels,
respectively; using the SAG2 parameter set) for the three different
approximations. The three models are in good agreement with observations in both
bands.
                                                                                
Changing the merger time-scales according to the new prescriptions of
J07 and B07 produces a decrease in the merger frequencies.
Both major and minor mergers are affected, although the changes are more important
for the minor mergers. This translates into a shift in the period of
maximum merger activity towards redshifts closer to $z \sim 1.5$, much lower than
that obtained for LC93, which occurs at $z \sim 2.5$. 
In turn, this affects the disc instability frequency
presented in Fig.~\ref{merTotDM} producing an order of magnitude increase in the number of events.
This is a result of the longer time-scales over which galaxies can form stars and 
accrete gas into the disc.
The higher number of disc instabilities and the delayed action of mergers
compensate each other, and no recalibration of the model parameters
is needed in order to obtain a good agreement between model and observational galaxy properties
such as the $z=0$ galaxy LF and 
bimodal galaxy colour distributions, stellar masses at different redshifts,
`BH-bulge' relations, and the SFH.
 
The new time-scales affect galaxy morphologies 
producing a 10 per cent excess of elliptical galaxies at intermediate stellar masses
with respect to observations (e.g. \citealt{Conselice06}).  
A possible reason for this slight excess of spheroids is that 
the frequency of disc instability events
increases producing a somewhat larger population of smaller elliptical galaxies than SAG2 and
the observations.
At the high mass end, the J07
prescription produces a slightly better agreement with the observed morphological fractions.

As it was mentioned above, the effect of the new merger time-scales
is difficult to detect in the galaxy statistics studied in this paper.
However, the study of quantities that are affected by the detailed stellar mass
growth may show more important differences, as would probably be the case for 
the amount of stellar mass formed in accreted satellites.

\section{Summary and Conclusions}\label{finalsec}

As several recent observational results indicate, AGN feedback
has become a very important element in our understanding on galaxy evolution 
 (e.g. \citealt{Nulsen07}, \citealt{Schawinski07}). In this paper, we study the 
effects of AGN feedback on the formation and evolution of galaxies by
using a combination of a cosmological {\em N}-body simulation of the concordance
$\Lambda$CDM cosmology and a semi-analytic model of galaxy
formation.  We have developed an AGN feedback model based
on different implementations considered in recent  semi-analytic models 
(\citealt{Croton06}, \citealt{Bower06} and \citealt{Malbon07}),
which are based on both, observations and a phenomenological picture of
the physics governing the extremely massive, central BHs, and of the gas ejected 
by these central engines back to the interstellar or even the intergalactic medium.
Our semi-analytic model is an improvement of the one described by
 \citet{Cora06} which involves gas cooling, star formation, galaxy mergers and
SN feedback, with the the advantage of including a proper treatment
of the chemical enrichment of baryons. The new version of this model (SAG2) 
benefits from the inclusion of
AGN feedback. In our model, the AGN activity is triggered by the accretion of
cold gas onto BHs, which
grow via three distinct channels: (i) mergers between two pre-existing black holes,
(ii) mass accretion during gas cooling processes, and (iii) gas accretion during starbursts. 
The latter are
produced during major and minor mergers, and disc instabilities,
being the last two processes new ingredients to our model. 
Black holes produce a luminosity which is proportional to the gas accretion
rate. 
When the accretion is produced during the gas
cooling process, the resulting AGN produces feedback (radio mode feedback). It
is also considered the possibility that AGN generated by the
BH growth during starbursts also produce feedback. 
Another novel feature analysed in 
our model is the influence of new prescriptions for merger time-scales on galaxy
properties.

We summarise the main results of this work.
\begin{itemize}
\item The growth history of galaxies is dominated by both major mergers and disc
instabilities, while minor mergers play only a minor role (Fig.~\ref{merTotDM}).
The frequency of starbursts peaks between $z\sim 2-3$ when produced by mergers, and between $z\sim 1$
when produced by disc instabilities.

\item The accretion history of black holes shows small variations 
between high and low mass BHs.
For high mass BHs at $z=0$, $M_{\rm BH}(z=0)>10^{8} M_{\odot}$,
the accretion is dominated by disc
instabilities at all redshifts.
Instead, mergers dominate at $z \gtrsim 2$ for lower masses, with
a larger influence of disc instabilites at lower redshifts (Fig.~\ref{BHgrow}).
Regardless
of the BH mass, the accretion via gas cooling processes is negligible.
Higher mass black holes assemble earlier in the history of the Universe
(Fig.~\ref{tracks}), in agreement with the downsizing scenario 
(e.g. \citealt{marconi04}, \citealt{shankar04}).               

\item We simultaneously obtain good
agreement with observations in several `BH-bulge' relations: 
BH-bulge mass (Fig.~\ref{BH-B}), BH mass as a function of bulge velocity dispersion (Fig.~\ref{BH-sigma}) and 
bulge luminosity (Fig.~\ref{BH-L}); these agreements were achieved using
a fixed value of the parameter $f_{\rm BH}$ (Eq.~\ref{massqsomode}) tuned to match 
the observed BH-bulge mass relation.

\item Our model predicts that a large fraction ($80\%$) of model QSOs need 
to be obscured
in order to reproduce the observed QSO LF (Fig.~\ref{LFQSO}).
Many studies have indicated this fact but fail to provide accurate
estimates of this
fraction (e.g. \citealt{gunn99}, \citealt{lacy07}). 
At low redshifts the model shows slightly
higher QSO number densities than the data, 
indicating the possibility that the obscured
fraction needs to become larger as the redshift decreases,
as suggested by some observational results
(e.g. \citealt{franceschini02}; \citealt{gandhi03}).

\item SAG2 reproduces many of the observed properties of the local galaxy
population that were also reproduced by the previous version of the model (e.g. the
Tully-Fisher relation, and the relation between gas fraction and absolute $M_{\rm
B}$ magnitudes). The observed bright-end of the galaxy
luminosity functions in the $b_{\rm J}$- and $K$-bands (Fig.~\ref{galLF}) have been
successfully recovered by SAG2 galaxies, being
one of the main successes of the
model. Our model also agrees very well with observational determinations of the
stellar mass function out to very high redshifts 
($z\sim 4.5$, Fig.~\ref{SMF}), as well as the global star
formation rate history (Fig.~\ref{SFRobserved}). The latter 
clearly demonstrates
how AGN feedback efficiently reduces star formation by quenching the gas cooling
process at low redshifts.

\item Our model reproduces the abundances of
observed morphological types \citep{Conselice06} at all stellar masses. 
We notice that the morphological fractions
are very sensitive to the AGN efficiency, which was set by comparing model
and observed galaxy LFs,
an independent statistic of the galaxy population. 
 
\item Our model galaxies show a bimodal {\em B-V} colour distribution up to stellar masses
$M_{\star} \approx 10^{10.5-11} M_{\odot}$. At higher masses, the distribution becomes
unimodal, presenting 
almost only red, elliptical galaxies. This constitutes one of the
main effects of AGN feedback since the original model shows a bimodal
distribution for all stellar masses with an important massive, blue spiral
population. 
The transition to a unimodal distribution is due to the passive evolution of
galaxies produced after exhausting their cold gas.

\item We have implemented two different channels of starburst feedback: `retention mode' and `ejection
mode' (indicating that gas is either retained or totally ejected from the galaxy halo,
respectively). We found that the ejection mode provides a good agreement
with all the observational constraints mentioned above. This mode does not require a
recalibration of model parameters with respect to those used in SAG2,
in contrast to the retention mode which
demands a higher AGN feedback efficiency ($\kappa_{\rm AGN}$) in order to reproduce the
observed galaxy LF; however, even after recalibrating the model, the 
morphological fractions in the retention mode
disagree with the observational measurements. This indicates that AGN feedback during
starbursts needs to be strong enough to expell the reheated gas
from the galaxy halo.
 
\item We implemented the new prescriptions for merger time-scales
presented by \citet{jiang07} and \citet{boylan07}
These prescriptions predict that major merger time-scales are significantly
shorter than those arising from the estimation of \citet{lacey93}. 
As result of this,
galaxies have more time to accrete gas and stars in the disc, thus producing
a significantly higher number of disc intability processes 
than what is shown in Fig.~\ref{merTotDM}.  
However, the net effect from applying
the new time-scales is difficult to detect, and
the $z=0$ galaxy properties (e.g. galaxy LF, Fig.~\ref{LFcomp}) 
remain almost unaffected.

\end{itemize}

Previous semi-analytic models were only focused on the analysis of particular properties
of galaxies and BHs. \citet{Croton06} present results on galaxy properties at z=0, 
while \citet{Bower06}, \citet{Cattaneo06} and \citet{Menci06} also take into account 
properties at high redshift but without a complete analysis of black
hole behaviour. On the other hand, \citet{Malbon07} make a complete analysis of BHs but
do not present an analysis on galaxy properties.
The combination of the different prescriptions of the black hole growth and the
associated AGN feedback taken from these theoretical works allowed us to develop a
model that is able to reproduce several observational constraints simultaneously, focusing
our analysis on different important results and predictions at low and high
redshifts. We obtain similar results for the galaxy LF and colour
distributions as previous models, but also obtain important differences in the evolution of the black
hole accretion rate; these differences may prove to be useful to obtain
more clues on the appropriate modelling and parametrization of starburst processes, chemical enrichment and
AGN feedback models.

We have presented a new model of galaxy formation and evolution
where we consider different mechanisms for the growth of black holes and galaxies, embedded in a
$\Lambda$CDM cosmology.  Among the several indicators that follow observational constraints, it has 
emerged that the black hole and bulge growth may seem `anti-hierarchical', in agreement with the
down-sizing scenario, even though the roots of their growth in our model is the hierarchical clustering
scenario. This phenomenon deserves a detailed analysis which we will tackle in a forthcoming paper.  
Further modifications to the model are being developed in order 
to improve the treatment of the black hole physics,
which involves the BH spin and different physical models of accretion (e.g.
\citealt{Bondi}, \citealt{Narayan97}) and
disc warping (e.g. \citealt{Shakura73}, \citealt{King05}). The
implementation of these processes will give us a
powerful tool to study into more detail the origin of the obscuration of QSOs and of the
radio-loud QSO population, the relation
between black hole properties and galaxy morphology, and several other interesting
observed AGN features. This will provide more links between
the diverse AGN phenomenon and new underlying still unknown processes involved in the
formation and evolution of galaxies.

\section{Acknowledgements}

We thank Richard Bower, Rowena Malbon, Carlton Baugh, Paulina Lira and
Mario Abadi for useful comments and
discussions, and Andrew Hopkins for providing the data compilation on
cosmic star formation rate density. We acknowledge the annonymous referee for
helpful remarks
that allow to
improve this work. CL acknowledges student support from Fondecyt
grant No. 1071006 and LENAC for support a visit to La Plata, Argentina; NP
was also supported by Fondecyt grant No. 1071006.  The
authors benefited from repeated visits of SC to Santiago de Chile supported by
Fondecyt grant No. 7070045.  This work was supported in part by the 
FONDAP Centro de Astrofísica, by
PIP 5000/2005 from Consejo Nacional de Investigaciones Cient\'{\i}ficas y T\'ecnicas,
Argentina, and PICT 26049 of Agencia de Promoci\'on Cient\'{\i}fica y T\'ecnica,
Argentina.

\label{lastpage}

\end{document}